\documentclass[twocolumn]{emulateapj}
\usepackage{epsfig}
\usepackage{amsmath}
\usepackage{graphicx}
\usepackage{natbib}
\usepackage[backref,breaklinks,colorlinks,citecolor=blue]{hyperref} 
\usepackage[all]{hypcap}     

\usepackage{footnote}
\usepackage{amssymb} 
\usepackage{color}
\usepackage{multirow}
\shorttitle{Stellar Masses and SFRs for 1M Galaxies from SDSS and WISE}
\shortauthors{Chang et al.}
\begin{document}
\title{Stellar Masses and Star Formation Rates for 1M Galaxies from
  SDSS$+$WISE}

\author{Yu-Yen~Chang\altaffilmark{1},
Arjen~van der Wel\altaffilmark{1},
Elisabete~da Cunha\altaffilmark{1,2},
Hans-Walter~Rix\altaffilmark{1}
}
\altaffiltext{1}{Max-Planck Institut f\"ur Astronomie, K\"onigstuhl 17, D-69117, Heidelberg, Germany; e-mail: \url{yu-yen.chang@cea.fr}} 
\altaffiltext{2}{Centre for Astrophysics and Supercomputing, Swinburne University of Technology, Hawthorn, Victoria 3122, Australia}

\begin{abstract}
We combine SDSS and WISE photometry for the full SDSS spectroscopic galaxy sample, creating SEDs that cover $\lambda$=0.4-22$\micron$ for an unprecedented large and comprehensive sample of 858,365 present-epoch galaxies. Using MAGPHYS we then model simultaneously and consistently both the attenuated stellar SED and the dust emission at 12$\micron$ and 22$\micron$, producing robust new calibrations for monochromatic mid-IR star formation rate proxies. These modeling results provide the first mid-IR-based view of the bi-modality in star formation activity among galaxies, exhibiting the sequence of star-forming galaxies (``main sequence'') with a slope of $d\log SFR/d\log M_*=$ 0.80 and a scatter of 0.39 dex. We find that these new star-formation rates along the SF main sequence are systematically lower by a factor of 1.4 than those derived from optical spectroscopy. We show that for most present-day galaxies the 0.4-22$\micron$ SED fits can exquisitely predict the fluxes measured by Herschel at much longer wavelengths. Our analysis also illustrates that the majorities of stars in the present-day universe is formed in luminous galaxies ($\sim L^*$) in and around the `green valley' of the color-luminosity plane. We make the matched photometry catalog and SED modeling results publicly available.
\end{abstract}
    
\section{Introduction}
\label{sec1}

Some of the most basic and important constraints on galaxy formation
models are the present-day stellar mass function and the distribution
of star formation among galaxies with different masses.  The Sloan
Digitital Sky Survey \citep[SDSS,][]{2005AJ....129.2562B,
  2008ApJS..175..297A, 2008ApJ...674.1217P} has provided the
measurements that underlie our current knowledge of the stellar masses
($M_*$) and star formation rates (SFRs) of large samples of galaxies
\citep[e.g.,][]{2004MNRAS.351.1151B}.
The SDSS multi-wavelength ($ugriz$) imaging has been used to estimate
luminosities and mass-to-light ratios
\citep[e.g.,][]{2012MNRAS.421..621B}. The resulting stellar masses
have been demonstrated to correlate tightly with total, dynamical mass
estimates \citep{2011MNRAS.418.1587T} with a scatter of only 0.13 dex.
\citet{2004MNRAS.351.1151B} use the \citet{2001MNRAS.323..887C}
photoionization model to convert nebular emission line fluxes from
SDSS spectroscopy into SFRs.  \citet{2007ApJS..173..267S} provided
SFRs based on GALEX UV fluxes.  
But these previous SDSS studies did not account for the extra information enclosed in dust emission, when estimating both star formation rate and dust attenuation.
Here we present alternative $M_*$ and SFR estimates by
extending the photometric wavelength coverage of the SDSS
spectroscopic sample to the near- and mid-infrared as enabled by WISE
\citep{2010AJ....140.1868W}, which provides allsky coverage at
$3-22\micron$.  

The $3.4\micron$ and $4.5\micron$ bands sample the Rayleigh-Jeans tail
of the stellar spectral energy distribution (SED), which avoids the
contribution from hot, young stars which can dominate at shorter
wavelenghts.  In addition, extinction is usually negligible in these
bands.  As a result, near-infrared luminosities provides fairly
accurate and precise stellar mass estimates, even in the absence of
any other photometric or spectroscopic information
\citep{2014ApJ...788..144M}.  Many authors have used $3-5\mu
m$-photometry from Spitzer or WISE to compare with stellar mass
estimates derived from photometry at $2\mu m$ and below
\citep[e.g.,][]{2007AJ....134.1315L, 2010RAA....10..329Z,
  2013AJ....145....6J}.  Here we combine SDSS and WISE data to
generalize the use $3-5\mu$m photometric information in estimating
stellar masses.

Mid-infrared emission traces star-formation activity through the
well-known correlation with PAH emission, sampled by the WISE
12$\micron$ band, and through the correlation with thermal radiation
from dust, sampled by the WISE 22$\micron$ band.  A number of authors
have compared star-formation rates from \citet{2004MNRAS.351.1151B},
based on optical emission lines, with mid-IR luminosities from WISE
\citep[e.g.,][]{2012ApJ...748...80D, 2013ApJ...774...62L,
  2013MNRAS.433.2946W, 2014MNRAS.438...97W}; these studies do not
model the mid-IR luminosity to obtain SFRs that are independent of the
emission-line based SFRs.

Here we take the important step to include the full WISE photometry
and employ SED modeling that consistently treats the stellar emission
along with the dust extinction {\it and} emission. This will result in
more robust masses and star-formation rates for dusty galaxies and, in
general, an alternative to the emission-line based SFRs from
\citet{2004MNRAS.351.1151B} for the full SDSS spectroscopic galaxy
sample.  This is not been attempted before for large samples of
galaxies.  \citet{2013AJ....145....6J} provided a detailed
multi-wavelength study of a small number of objects and will provide
an important benchmark to test our results.
\citet{2014ApJS..212...18B} performed SED fits across the UV to mid-IR
wavelength in an exercise similar in approach as what we present here,
but for those authors the focus lay on producing a set of
representative templates: stellar mass and SFR estimates are not
presented or discussed.  \citet{2014ApJ...782...90C} analyze the
mid-IR properties of the large GAMA sample, but there H$\alpha$-based
SFRs were used as a calibrator for the mid-IR luminosities from WISE.
As such, our study is the first to use the mid-IR luminosities of a
large sample of galaxies (the full SDSS spectroscopic galaxy sample)
to estimate their SFRs in a manner that is entirely independent of
optical emission line luminosities and other external calibrators.

This paper is organized as follows. In Section 2 we will describe the
SDSS and WISE photometric datasets, with a particular focus on total
flux measurements for extended sources.  In Section 3 we will describe
MAGPHYS \citep{2008MNRAS.388.1595D}, the SED fitting code, and test
the robustness of the fitting results. In particular, we will test how
well MAGPHYS can predict, with the available wavelength coverage of
$0.4-22\micron$, the total IR luminosity.  Furthermore, we will
compare our $M_*$ and SFR estimates with the \citet{2004MNRAS.351.1151B} measurements.  In
Section 4 we will describe the publicly released catalog with the
fitting results for the entire SDSS spectroscopic sample
($\sim$800,000 galaxies).  In Section 5 we will show, as an
illustration, how star formation is distributed over galaxies with
different masses and colors.

We use AB magnitudes and adopt the cosmological parameters
($\Omega_M$,$\Omega_\Lambda$,$h$)=(0.30,0.70,0.70) and adopt the
\citet{2003PASP..115..763C} stellar initial mass function.

\section{Data}
\label{sec2}

\begin{figure}
\centering
\includegraphics[width=1.0\columnwidth]{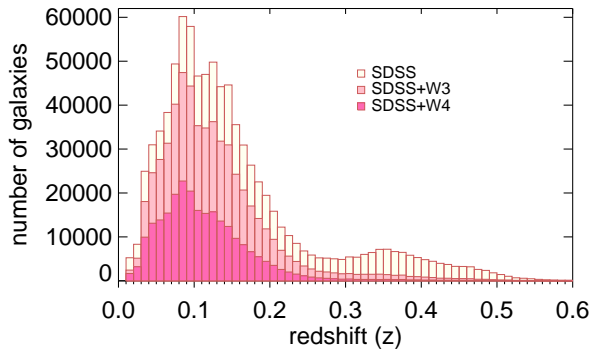} 
\caption[]{Redshift distribution of the SDSS galaxy sample.  98.24\%
  of the objects are detected in at least one WISE band.  The
  subsamples with $>2\sigma$ detections in W3 (12$\mu$m) and W4
  (22$\mu$m) are highlighted. }
\label{swfig_z}
\end{figure}

\label{sample}
We use the SDSS spectroscopic galaxy sample as compiled in the New
York University Value-Added Galaxy Catalog
\citep[NYU-VAGC,][]{2005AJ....129.2562B, 2008ApJS..175..297A,
  2008ApJ...674.1217P} \footnote{\url{http://sdss.physics.nyu.edu/vagc/}},
which contains 858,365 galaxies with reliable redshift measurements
({\tt OBJTYPE}=GALAXY, {\tt Zwarning}=0), distributed as shown in
Figure~\ref{swfig_z}.  We adopt {\tt MODELFLUX} and {\tt
  MODELFLUX\_IVAR} as the flux measurements and their uncertainties,
with galactic extinction corrections as prescribed by
\citet{2011ApJ...737..103S}.  To account for the uncertainty in the
extinction law, we propagate an uncertainty in $R_V=3.1 \pm 0.2$ in
our flux uncertainty estimates.  Minimum uncertainties are added in
quadrature: 0.05 mag for the $u$ band and 0.02 mag for the $griz$
bands.  This is aimed at preventing from small systematic
uncertainties to dominate the our goodness-of-fit assessment and the
uncertainties in our derived parameters.
In case of catastrophic failures in the photometric measurements, which occasionally occurs in the $u$ and $z$ bands, we omit these data points from the fits.

The SDSS parent sample is cross-matched with the AllWISE source
catalog\footnote{\url{http://wise2.ipac.caltech.edu/docs/release/allwise/}}.
The vast majority of galaxies have counterparts (98.24\%),
identified as the brightest object within a search radius of 6 arcsec, similar to
the WISE PSF. 
In line with the approach taken by \citet{2012ApJ...748...80D} and \cite{2013AJ....145...55Y}, most matches (99.08\%) are found within a radius of 3 arcsec and the majority (91.20\%) do not have multiple matches within 6 arcsec. We provide a flag ($FLAG\_W$) that indicates if there are one or more matches, but we do not exclude any matched sources {\it a priori}.
Various flux measurements are provided by the AllWISE
catalog.  In order to minimize the effects of source blending we use
the {\tt W?mpro} and {\tt W?sigmpro} flux measurements and their
uncertainties.  These are profile-fitted photometry measurements,
performed simultaneously on neighboring sources, using the point
spread function (PSF) as the source model.  Most of the galaxies with
counterparts have significant W1 and W2 flux measurements (98.83\%).
For W3 and W4, many of the flux measurements amount to upper limits
(29.10\% and 70.89\%, respectively).  The redshift distributions of galaxies
with $>2\sigma$~W3 and W4 detections are shown in Figure \ref{swfig_z}.
We will provide flags to indicate which galaxies have detections, and
which have upper limits.  The usefulness of the upper limits will
depend on the goals of the user.

Galaxies are typically smaller than the WISE PSF, but the amount by
which {\tt W?mpro} underestimate the total flux due to the spatial
extent of the sources is not necessarily negligible.  The W1, W2, and
W3 bands all have very similar PSFs with FWHM$\sim6-7$~arcsec, while
the W4 PSF is larger ($\sim12$)~arcsec.  We investigate the missing
light fraction for W1, W2, and W3 by generating with {\tt galfit}
\citep{2010AJ....139.2097P} a series of simulated light distributions with
2-dimensional S\'ersic profiles convolved with the W1 PSF models
provided by \citet{2011PASP..123.1218A}, and inserted into empty
sections of real W1
images\footnote{\url{http://www.astro.princeton.edu/~ganiano/Kernels/Ker\_2012\_May/PSF\_fits\_Files/}}.
Then, we use {\tt galfit} to fit the PSF model to the simulated images
to measure the PSF profile flux.  We find that the difference between
the PSF profile flux and the true, total flux is mostly a function of
the effective radius of the S\'ersic profile, and hardly depends on
input flux, axis ratio, or S\'ersic index. 
For a typical size of 5 arcsec, we find that $\Delta m$ varies from 15.63 mag for S\'ersic index $n=1$ to 15.65 mag for Sersic index $n=4$.
Based on these simulations
we find that the PSF profile magnitude underestimates the total flux
by
\begin{equation}
  \Delta m  = 0.10+0.46\log(R_{e})+0.47\log(R_{e})^2+0.08\log(R_{e})^3,
\label{dm}
\end{equation} 
where $R_{e}$ is the effective radius in arcsec.  This correction is
used whenever the size is larger than $R_{e}=0.5$~arcsec.
We propagate the uncertainty in the adopted radius into Delta m, which is in turn propagated into the flux uncertainty in flux.

We use the $r$-band effective radii and apply a systematic correction of a factor $1.5\pm 0.2$ downward to to convert from optical size to near-IR size. This conversion is based on the analysis of \citep{2014MNRAS.441.1340V}, who consistently measured sizes for a large set of galaxies across the wavelength range 0.4 to 2.2 $\micron$. These
corrected sizes and Equation \ref{dm} are then used to correct the W1, W2, and W3 fluxes.
We use $r$-band effective radii are measured by \citet{2011ApJS..196...11S} when available, and otherwise {\tt deVRad} or {\tt expRad from} the SDSS catalog as appropriate. We prefer the \citet{2011ApJS..196...11S} measurements as their fitting methodology (two-dimensional light profiles) is more similar to the methodology used by \citet{2014MNRAS.441.1340V} (the SDSS
pipeline fits one-dimensional light profiles). We note that the difference between the two versions of our W1/2/3 magnitude corrections is small: for objects for which both measurements are available the median difference is 0.03 mag, and the random
scatter 0.14 mag.  These uncertainties are small compared to the uncertainties in stellar masses and SFRs derived below.
These final flux measurements, along with the applied corrections, are
listed in the public catalog (see Section \ref{sec4}).  
The median correction is 0.25 mag, with a scatter of -0.07 (16\%-ile) and +0.18 (84\%-ile).
Corrections for W4, for which the PSF is twice as wide, are not made.

To account for possible systematic uncertainties of the WISE and SDSS
photometric systems and further systematic differences between the
total flux measurements we adopt and propagate 0.1 mag uncertainties
for all WISE fluxes.


\section{SED Modeling}
\subsection{Method}
\label{sec3}

We use MAGPHYS to fit the photometric SED
\citep{2008MNRAS.388.1595D,2012IAUS..284..292D}
\footnote{\url{http://www.iap.fr/magphys}}.  The public version of
MAGPHYS contains 50,000 stellar population template spectra (the
optical photometric library) and 50,000 PAH+dust emission template
spectra (the infrared photometric library).  The stellar emission
templates use the \citet{2003MNRAS.344.1000B} stellar population
synthesis models and are generated for a wide range of star formation
histories parameterized as exponentially-declining models with
superimposed random bursts of star formation.

The SEDs are computed by adding the individual spectra
of all simple stellar populations, weighed by mass.  The original
MAGPHYS library was constructed for modeling IR-luminous,
star-bursting galaxies. The original MAGPHYS library from the public version with 50,000 stellar population template spectra was constructed for modelling star-forming galaxies and we extended the optical
library to include 25,000 additional templates for more passive
stellar populations.

The infrared templates describe emission by dust.  The total dust
luminosity over 3 to 1000$\micron$ has components of emission from
PAHs and dust with a range of temperatures.  The model contains the
ambient (diffuse) interstellar medium and star-forming regions (birth
clouds).  Because stars are born in dense molecular clouds which
dissipate typically after $10^7$ years, the SED of young stellar
populations are attenuated by both dust in the birth clouds and dust
in the ambient ISM, while the SED of older populations is only
attenuated by dust in the diffuse ISM.  The absorbed light is assumed
to be re-emitted in the mid- and far-infrared, requiring conservation
of energy.

MAGPHYS fits SEDs in the observed frame.  Therefore, we generate
libraries with model fluxes in the observed frame for a series of
narrow redshift bins ($\delta z=0.0001$) that span the range of our sample.
The fitting process for each individual galaxy is then expediated in
two ways. First, we only consider optical templates with similar
$(g-i)_{\rm{model}}$ colors as the observed $(g-i)_{\rm{data}}$ color.
To be precise, we select templates with $|(g-i)_{model}-(g-i)_{data}|<
0.05 + \sigma[(g-i)_{data}]$, where $\sigma$ refers to the
uncertainty. 
This method works for any two filters, and here we choose $g-i$.  The selection is inclusive enough to avoid changes in the results: none of the models eliminated from consideration have significant likelihood values. We check that we would get similar results if we fitted the whole model library.
 Second, we draw a random set of 1,000 infrared templates
from the total set of 50,000 templates.  Since our data do not sample
the thermal peak, we cannot stringently constrain the dust temperature
distribution, rendering full exploration of parameter space useless.
As a test we compare the results based on the full and reduced infrared libraries and find no systematic differences (smaller than 0.01 dex for both stellar mass and SFR) 
in the fitting results and no increases or decreases in the formal fitting
uncertainties.

In a small number of cases we find that MAGPHYS allows the presence of
an unrealistically large amount of cold dust.  In order to avoid this,
we put a mild constraint on the model 250$\mu$m luminosity based on
the maximum observed 250$\mu$m-to-22$\mu$m flux ratios
\citep{2001ApJ...556..562C}: $L\nu_{250\micron }$[$L_\odot/Hz$] $<$
85.7 $\times (L\nu_{22\micron}+\sigma (L\nu_{22\micron}))$
[$L_\odot/Hz$], which is $\sim$6 times the typical ratio.  We will
investigate the precision of our predicted total IR luminosities
below, in \S~\ref{hatlas}.

Given our 10 flux constraints (SDSS$ugriz$; W1$-$4;
$L\nu_{250\micron}$ limit) $\chi^2$ is calculated for each model in
the library.  
For each model parameter the posterior probability distribution can be generated by taking the prior probability distribution (from the full library) and assign weights $exp(-\chi^2/2)$  to each of the models. In doing so we marginalize over all other model
parameters and assume that the likelihood of a given model given the data is proportional to $exp(-\chi^2/2)$.
We adopt the 50 percentile value -- that is, the median -- as the best
estimate. In practice, in the case of upper limits, we assumed zero flux with the upper limit as the error bar.

Rest-frame fluxes are calculated using the observed fluxes and colors.
Analogous to \citet{2012ApJ...749...96H} we derive for each redshift
$z$ bin a linear fit between, on the one hand, observed-frame
magnitudes ($m_{obs1}$, $m_{obs2}$) of the templates in the filters
straddling the desired rest-frame band, and, on the other hand, the
rest-frame magnitude ($m_0$) of the same templates: $m_0$ = $m_{obs1}$
+ $A(z)$ $\times$ ($m_{obs1}$ $-$ $m_{obs2}$ ) + $B(z)$.  Then, for
each galaxy in the sample, we use the observed magnitudes and the
values of $A(z)$ and $B(z)$ to compute its rest-frame magnitudes.  We
also adopt this technique to calculate rest-frame 12$\mu$m and
22$\mu$m luminosities.  The basic result of the SED MAGPHYS modelling
for each object, is a joint PDF for the 16 parameters,
which we then characterize by the median and the percentiles of the
marginalized distributions.

\begin{figure}[h]
\centering
\caption[]{Rest-frame $u-r$~vs.~$r-z$ distribution for our SDSS+WISE
  sample, color-coded according to the fraction of objects that lie on
  the star-forming sequence or significantly below according to their
  full SED fit (see Section \ref{sec4}).  The clearly distinct
  color-color distributions of star-forming and quiescent galaxies
  validate the assumption that a color-color diagram can be
  effectively used to distinguish the two types of galaxies, as is
  often done at high redshift.  The labels correspond to the objects
  for which the SEDs are shown in Figures \ref{massive}, \ref{dusty},
  \ref{blue}, and \ref{swfig_hatlas}.}
\includegraphics[width=1.00\columnwidth]{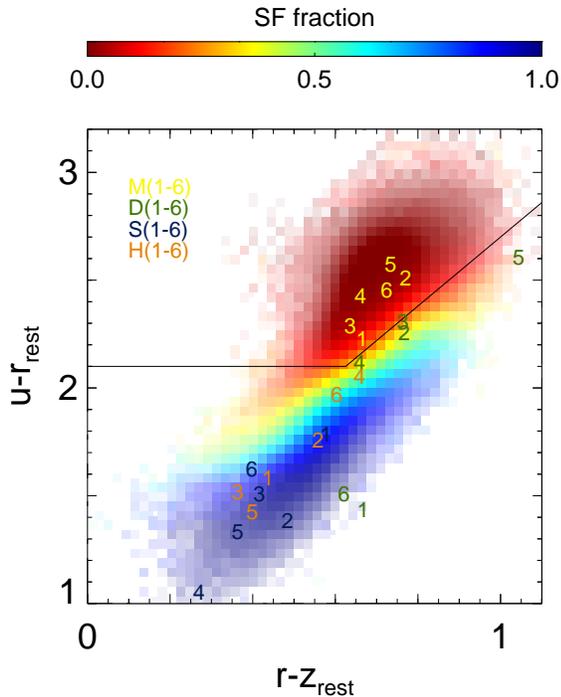}
\label{swfig_urz}
\end{figure}

\begin{figure*}
\centering
\includegraphics[width=1\columnwidth]{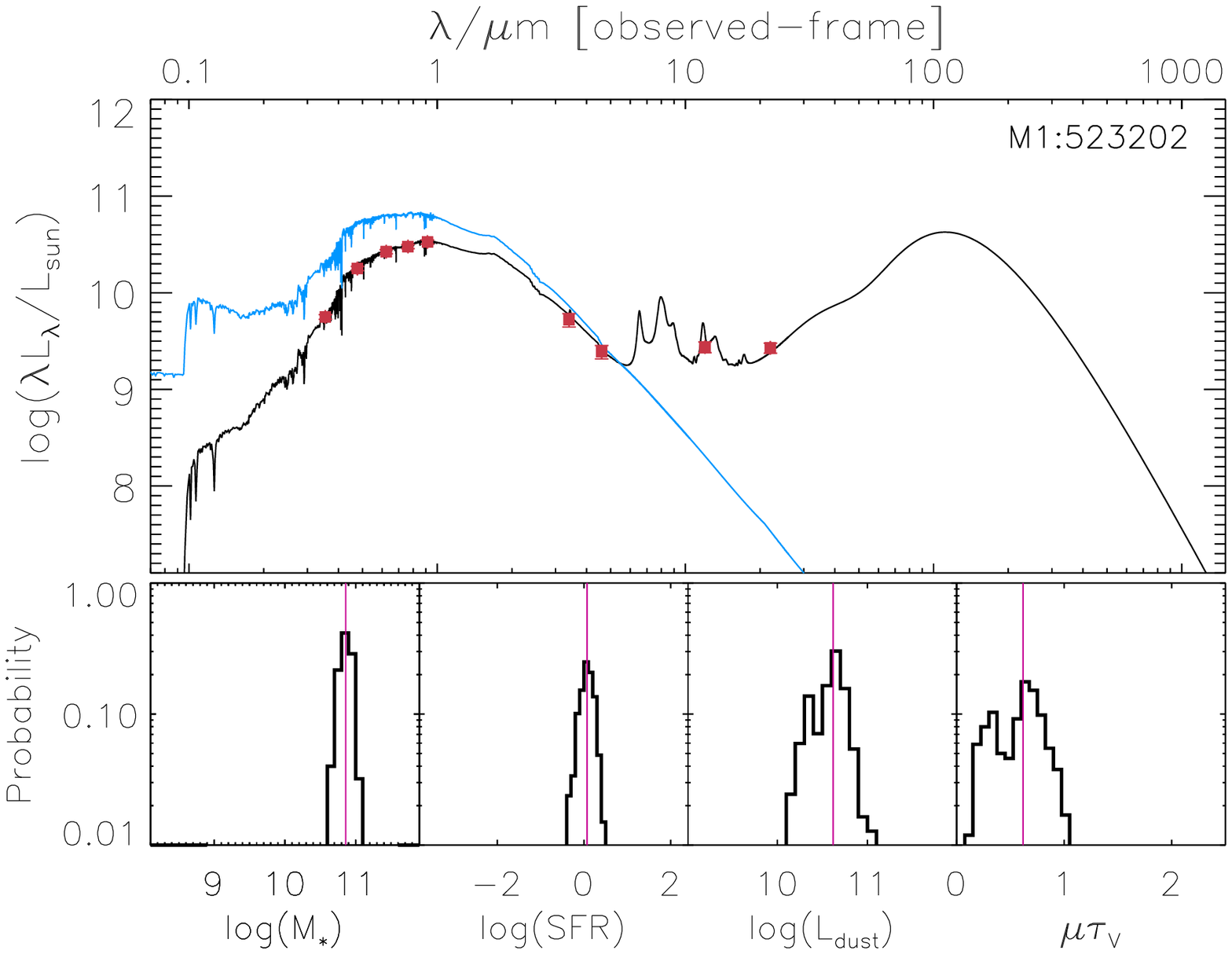}
\includegraphics[width=1\columnwidth]{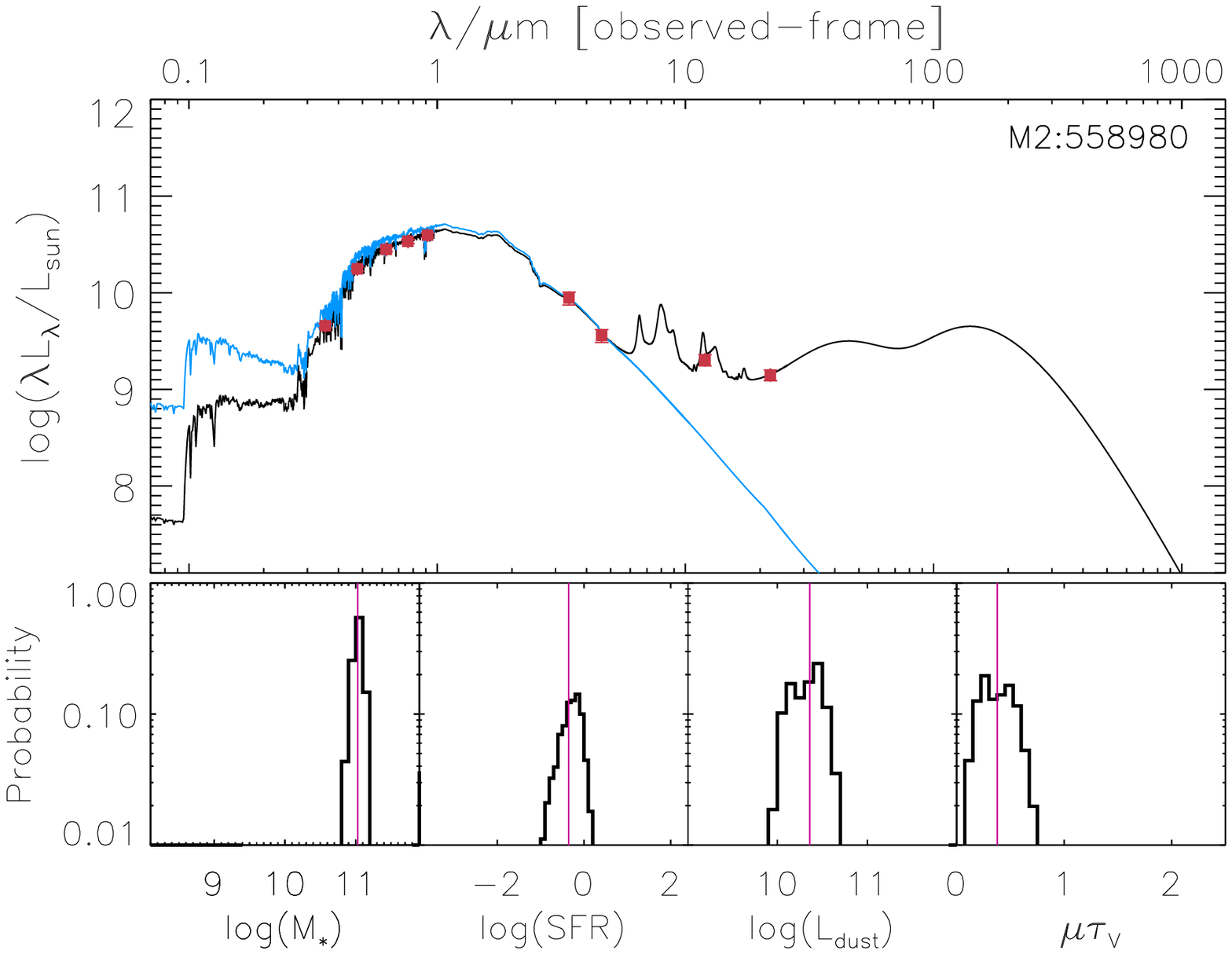}
\includegraphics[width=1\columnwidth]{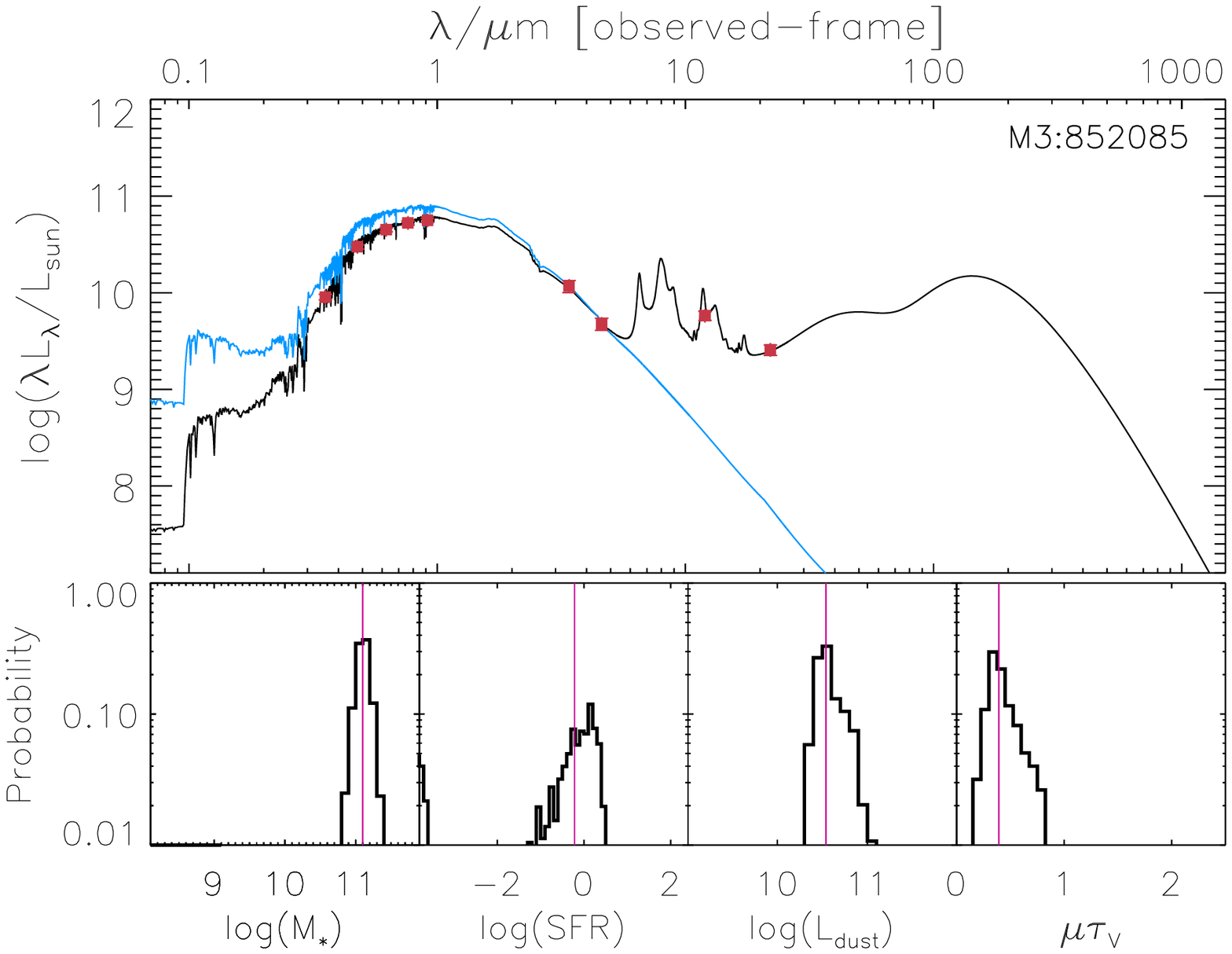}
\includegraphics[width=1\columnwidth]{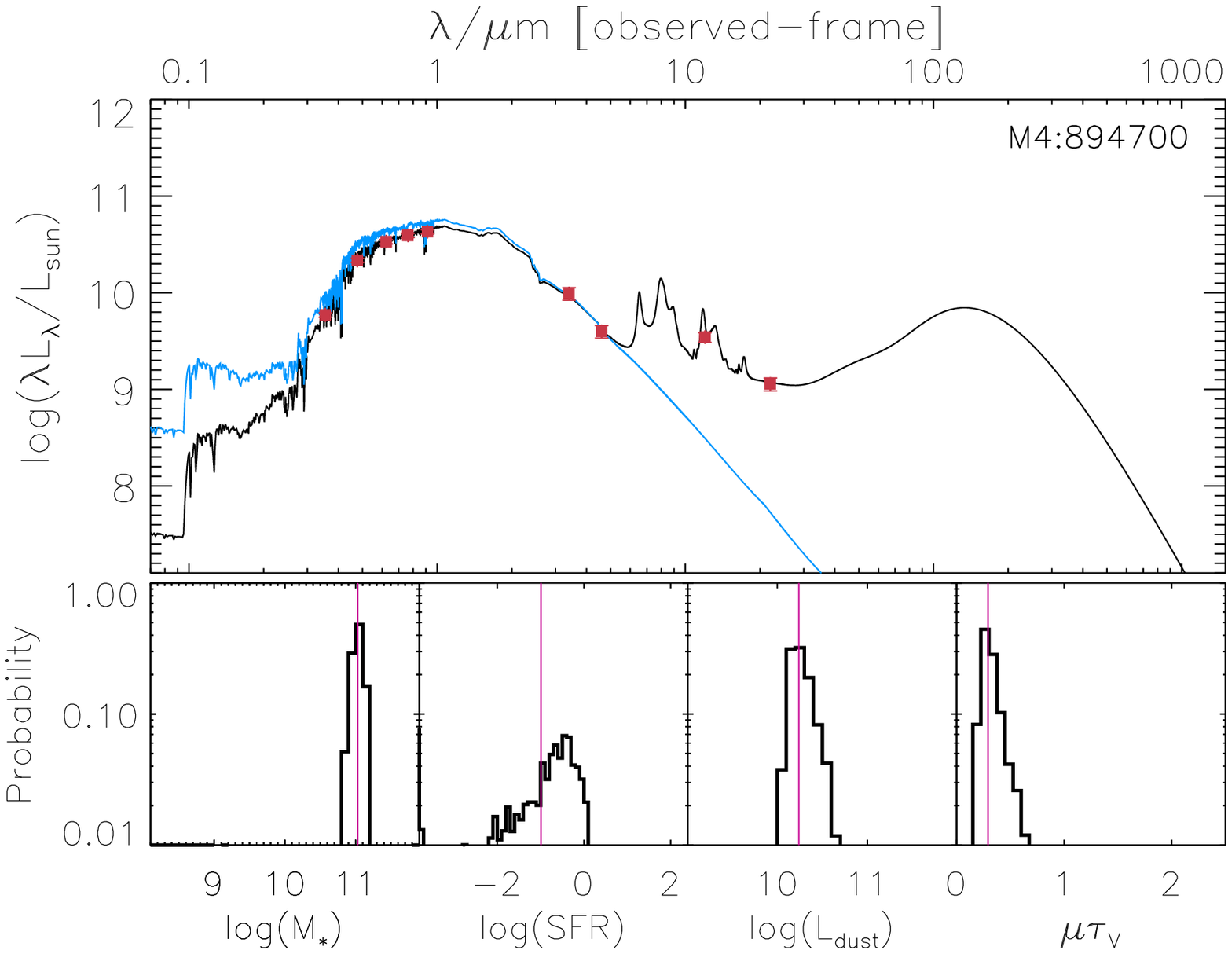}
\includegraphics[width=1\columnwidth]{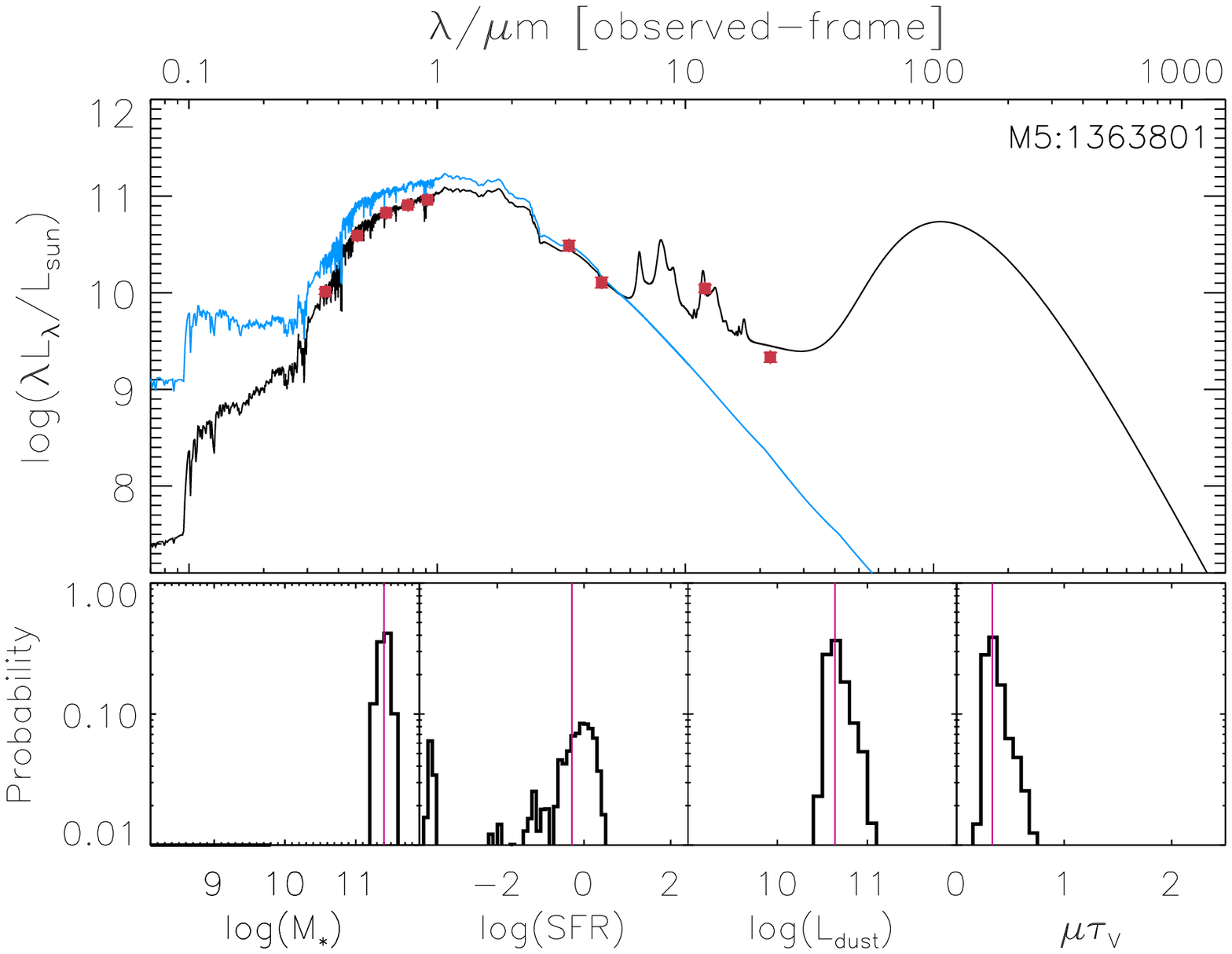}
\includegraphics[width=1\columnwidth]{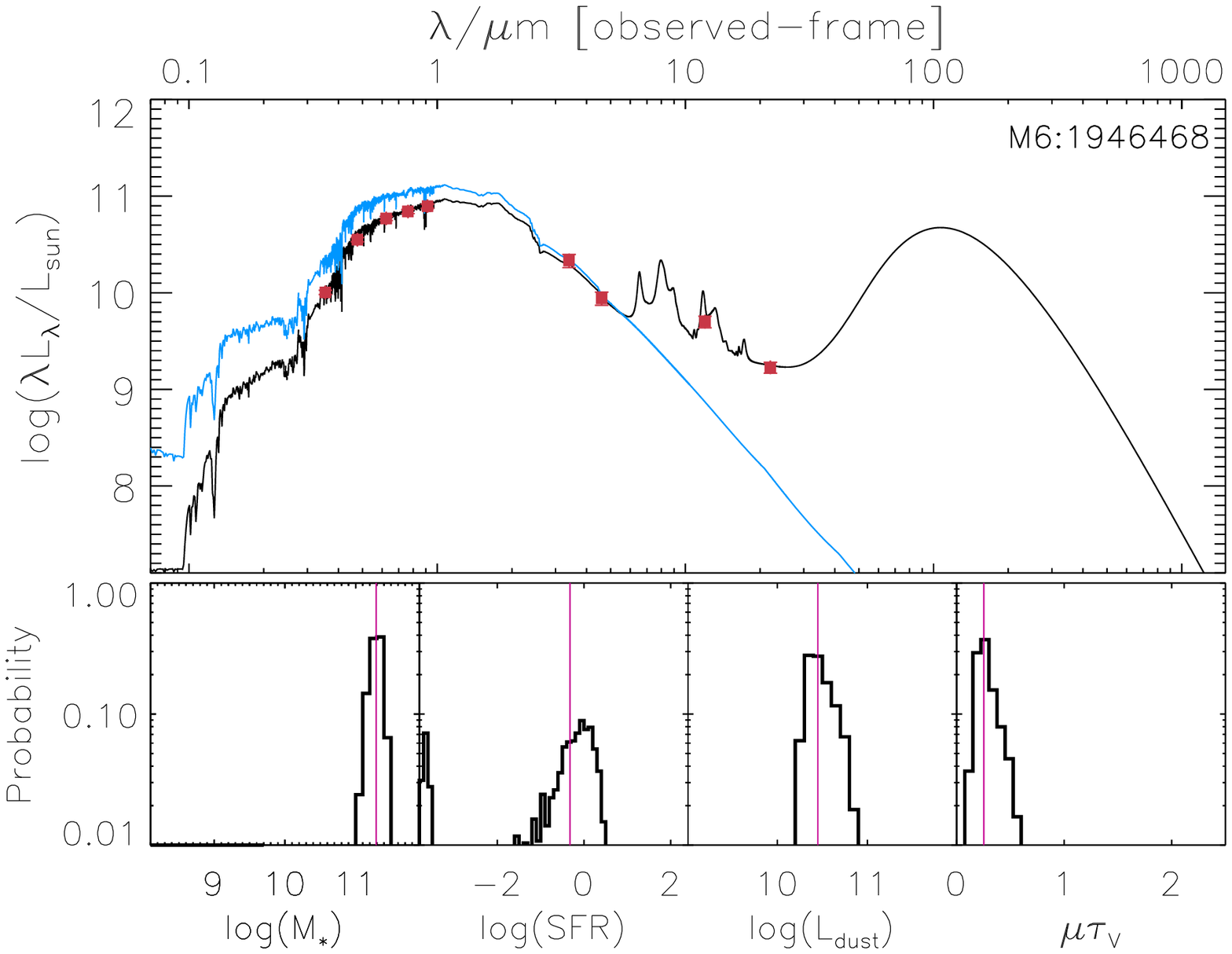}
\caption[]{Example SEDs and fitting results of six massive, quiescent
  galaxies.  The large sub-panels show the observed fluxes (red
  points), the best-fitting SEDs (black lines) and the corresponding
  unattenuated stellar SEDs (blue lines). The smaller panels show the
  marginalized posterior probability distributions of four models
  parameters: $M_*$, SFR, dust luminosity (in solar units), and dust
  attenuation.  The red lines show the median values, which we adopt
  as best-fitting values of the PDF.  The short-hand IDs (M1-6)
  correspond to the labels in Figure \ref{swfig_urz}.}
\label{massive}
\end{figure*}

\begin{figure*}
\centering
\includegraphics[width=1\columnwidth]{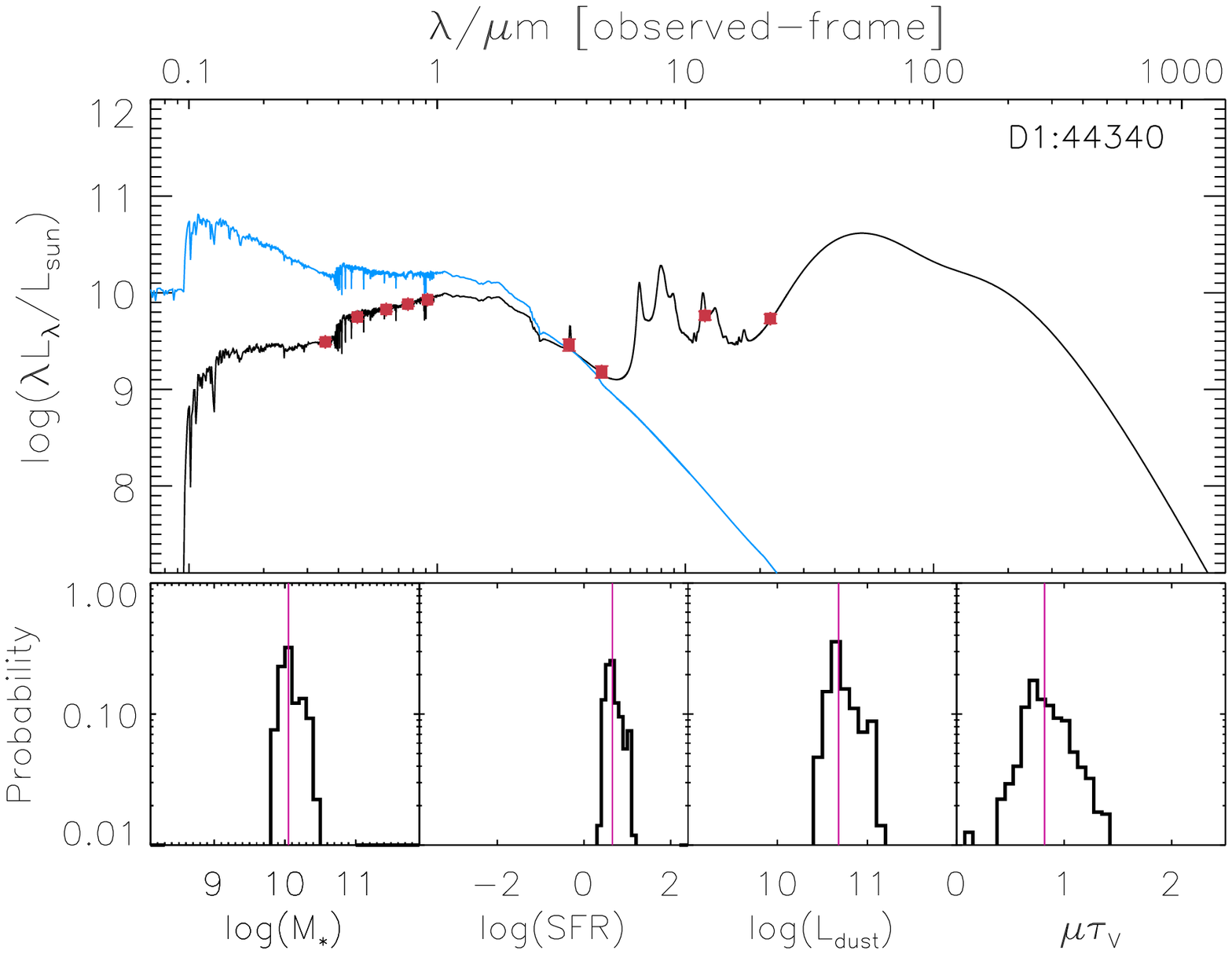}
\includegraphics[width=1\columnwidth]{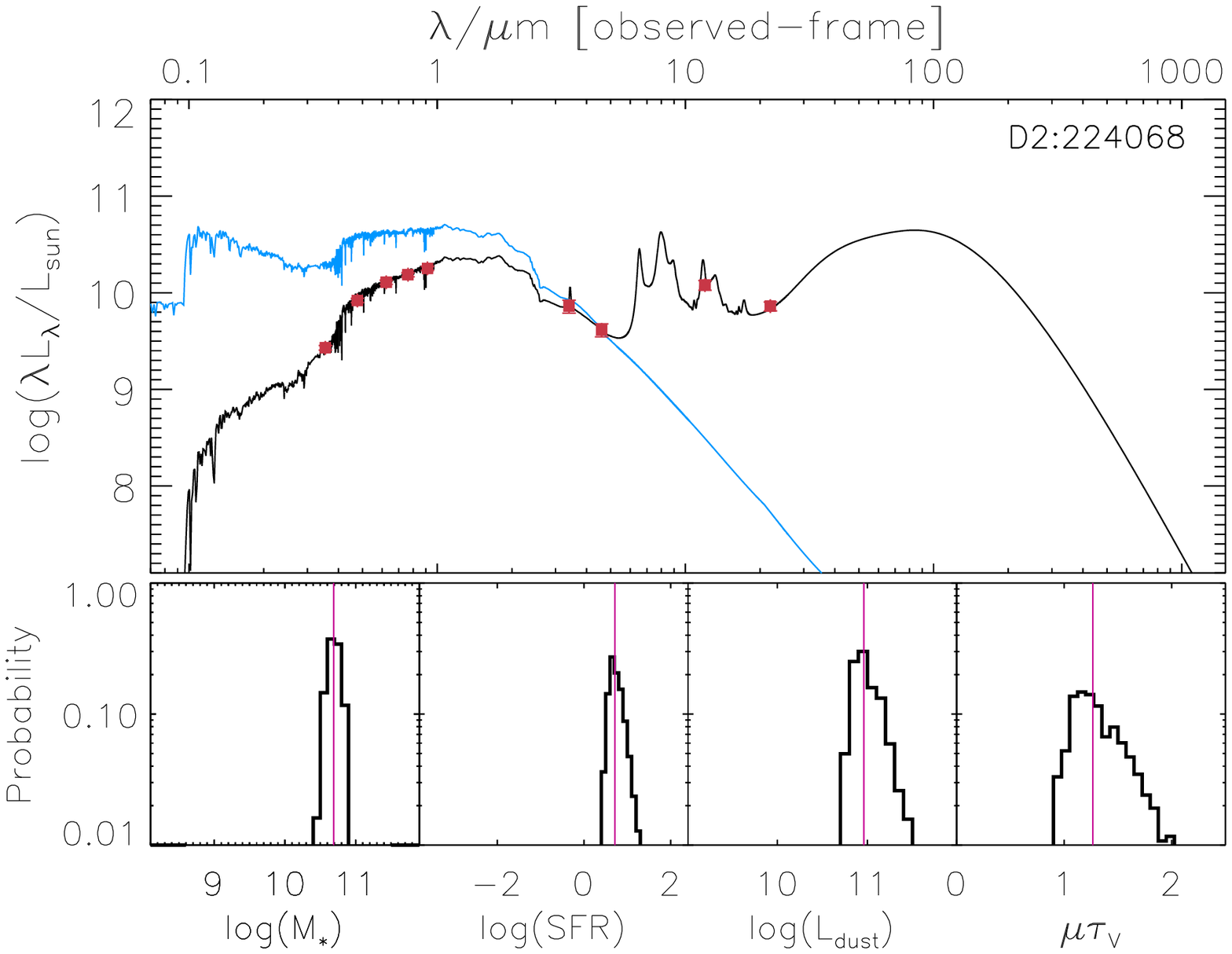}
\includegraphics[width=1\columnwidth]{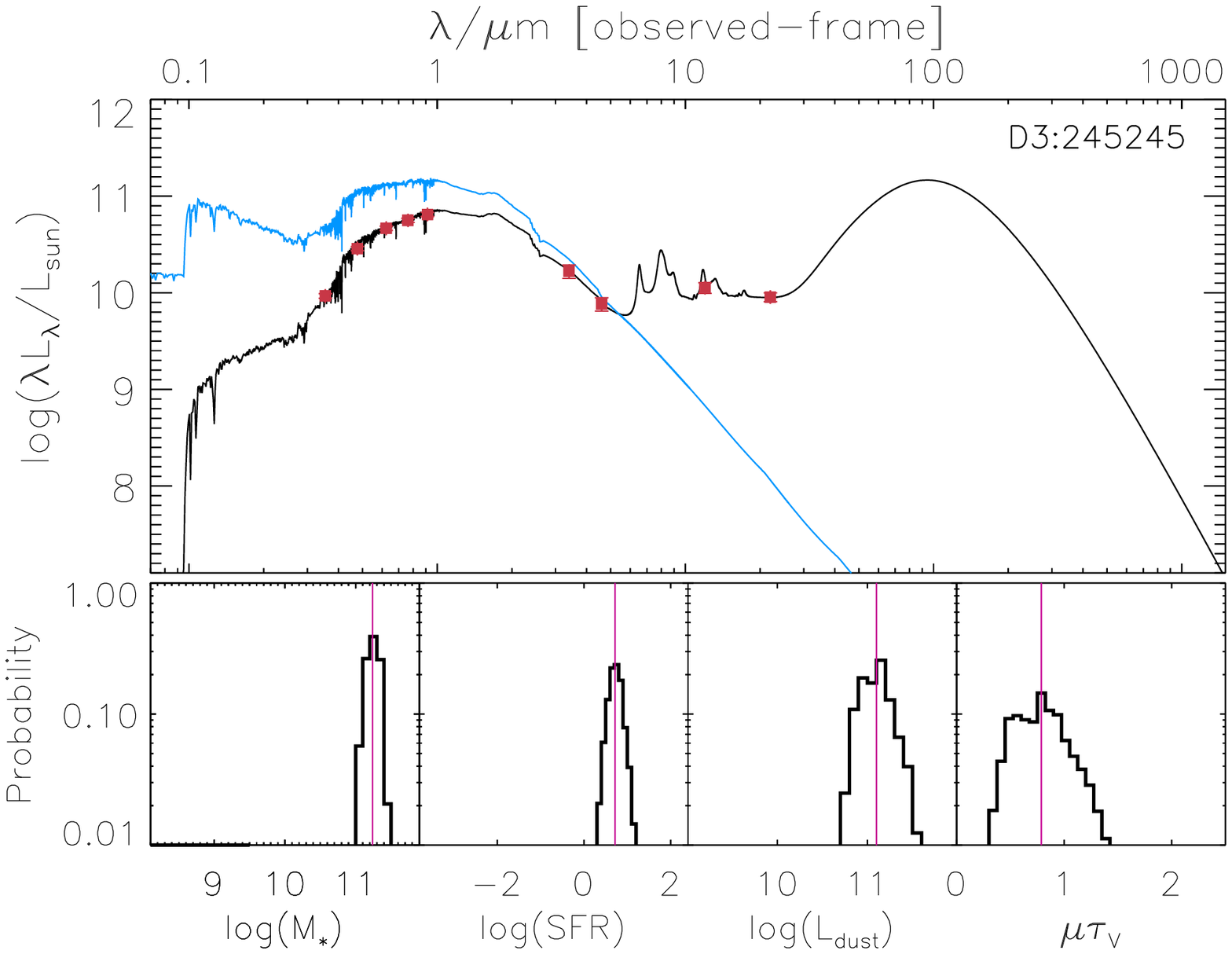}
\includegraphics[width=1\columnwidth]{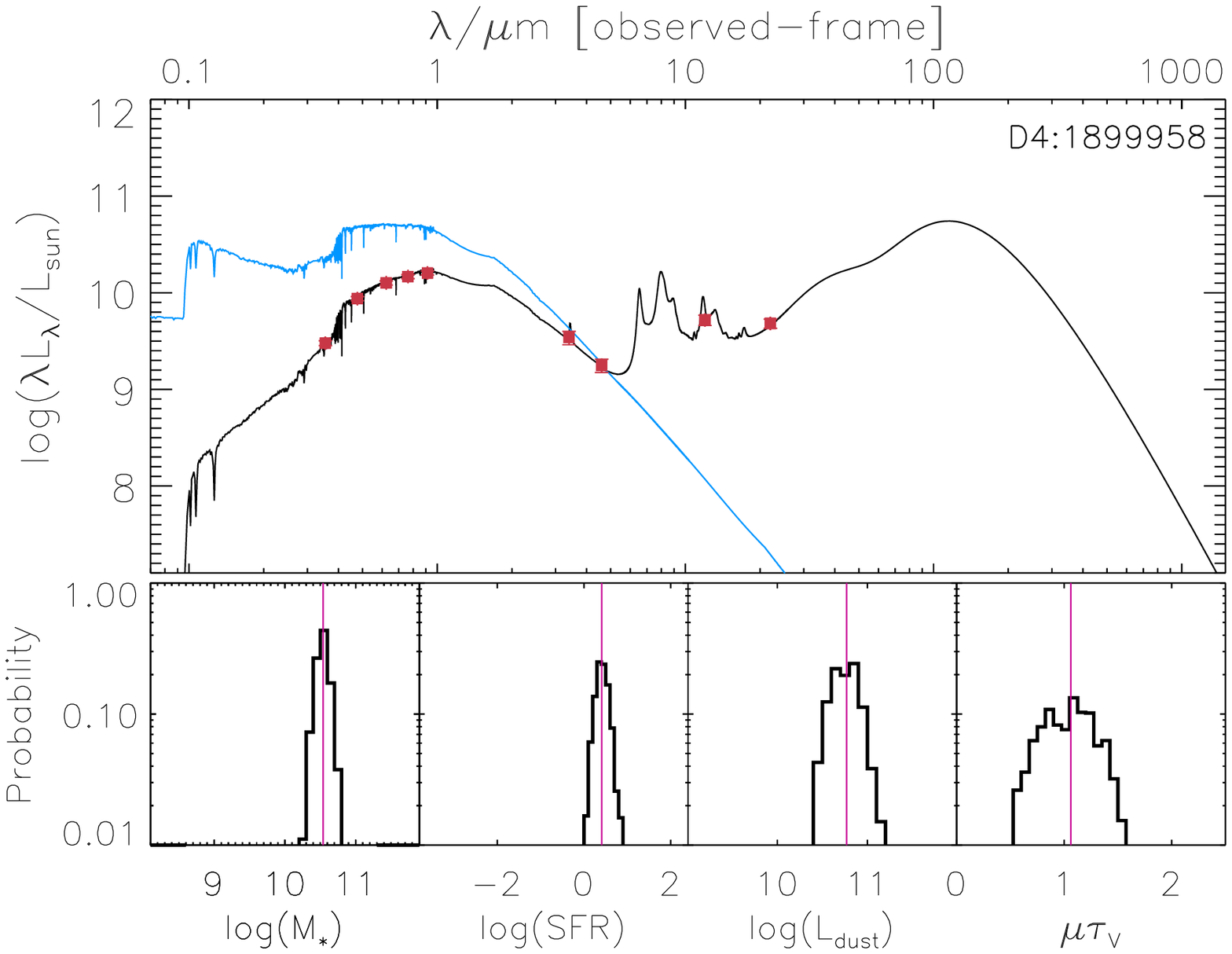}
\includegraphics[width=1\columnwidth]{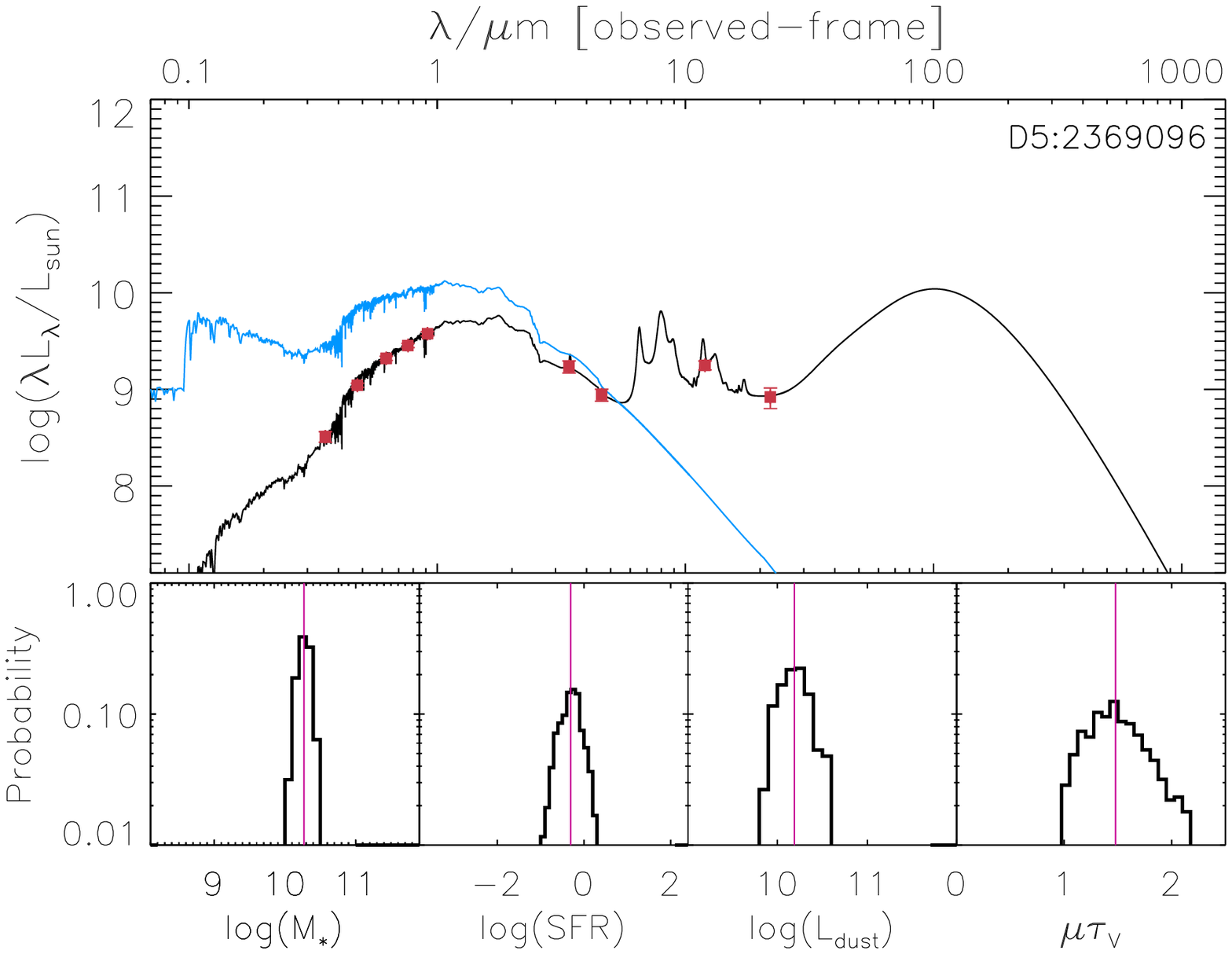}
\includegraphics[width=1\columnwidth]{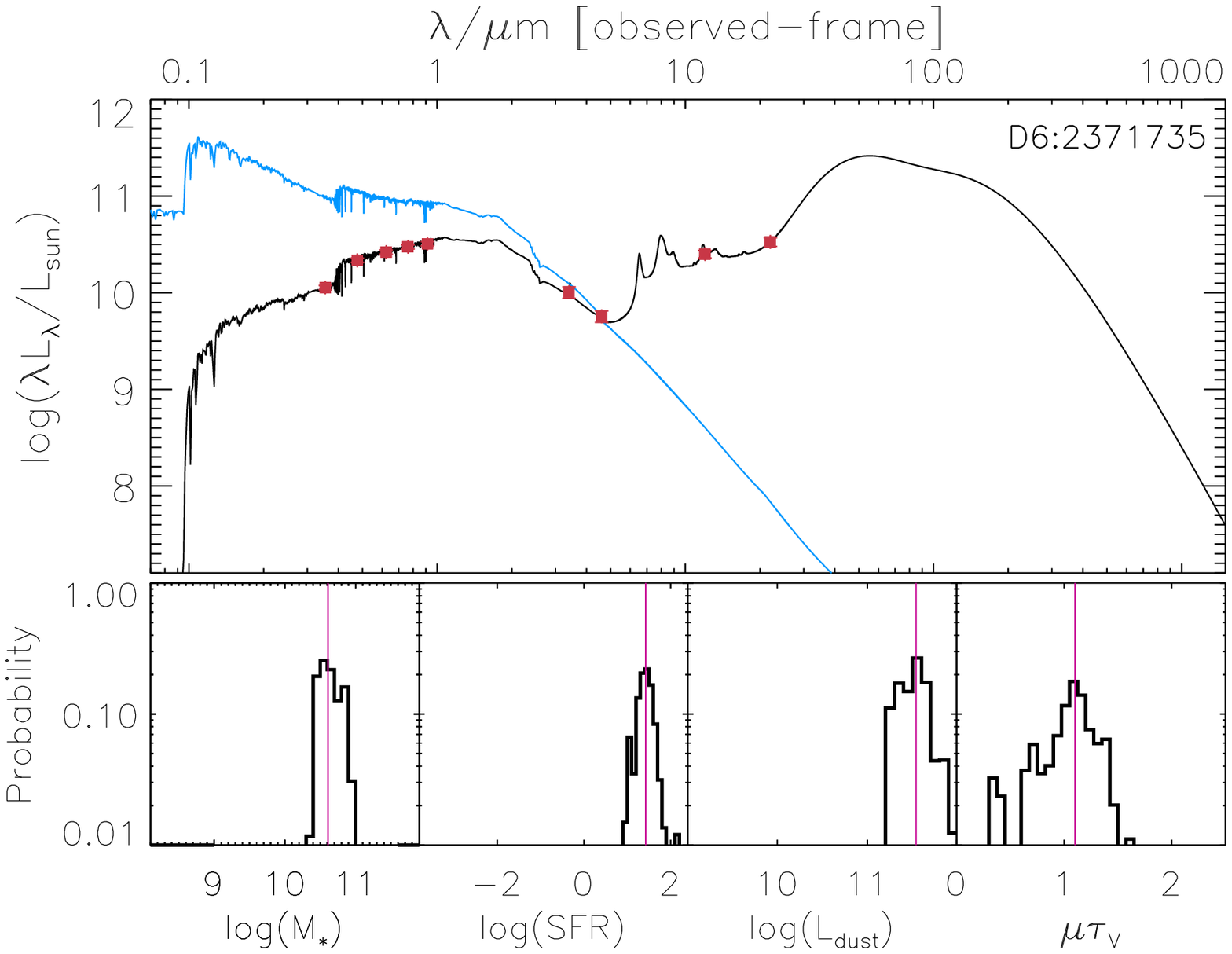}
\caption[]{Example SEDs and fitting results of six star-forming
  galaxies , where the optical light is significantly dust-extincted.
  See Figure \ref{massive} for an explanation of the
  panels, lines, and symbols. The short-hand IDs (D1-6) correspond to
  the labels in Figure \ref{swfig_urz}. }
\label{dusty}
\end{figure*}

\begin{figure*}
\centering
\includegraphics[width=1\columnwidth]{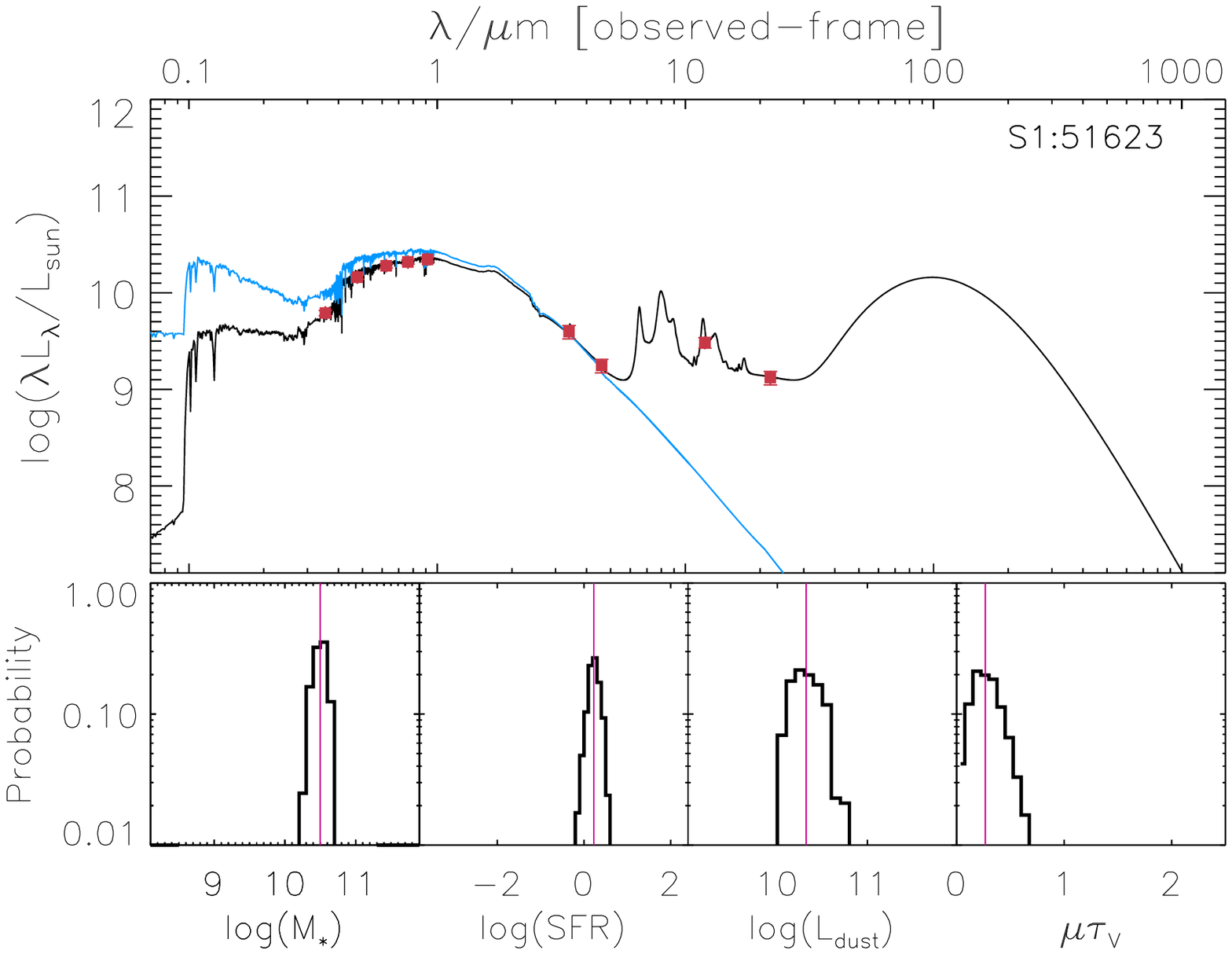}
\includegraphics[width=1\columnwidth]{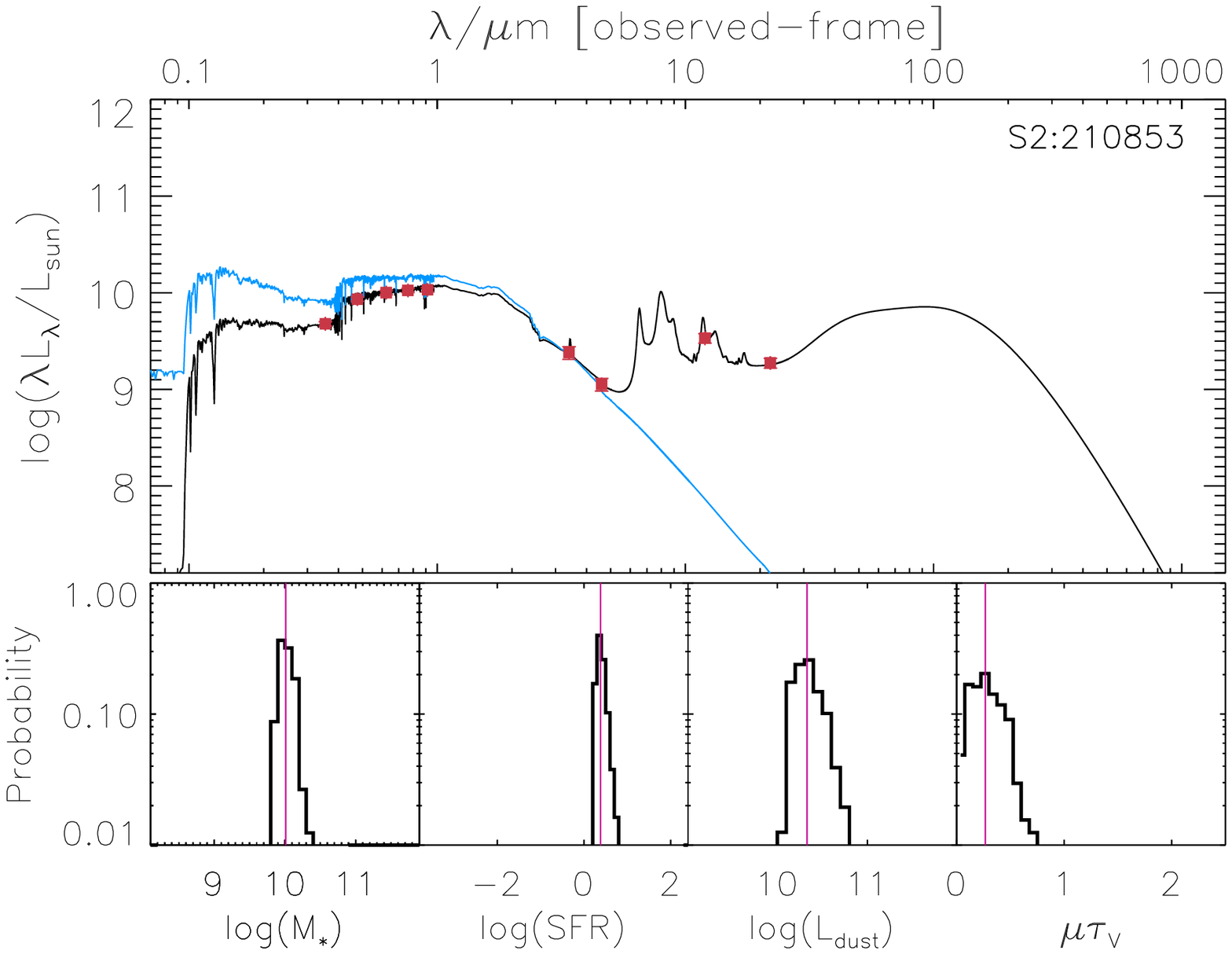}
\includegraphics[width=1\columnwidth]{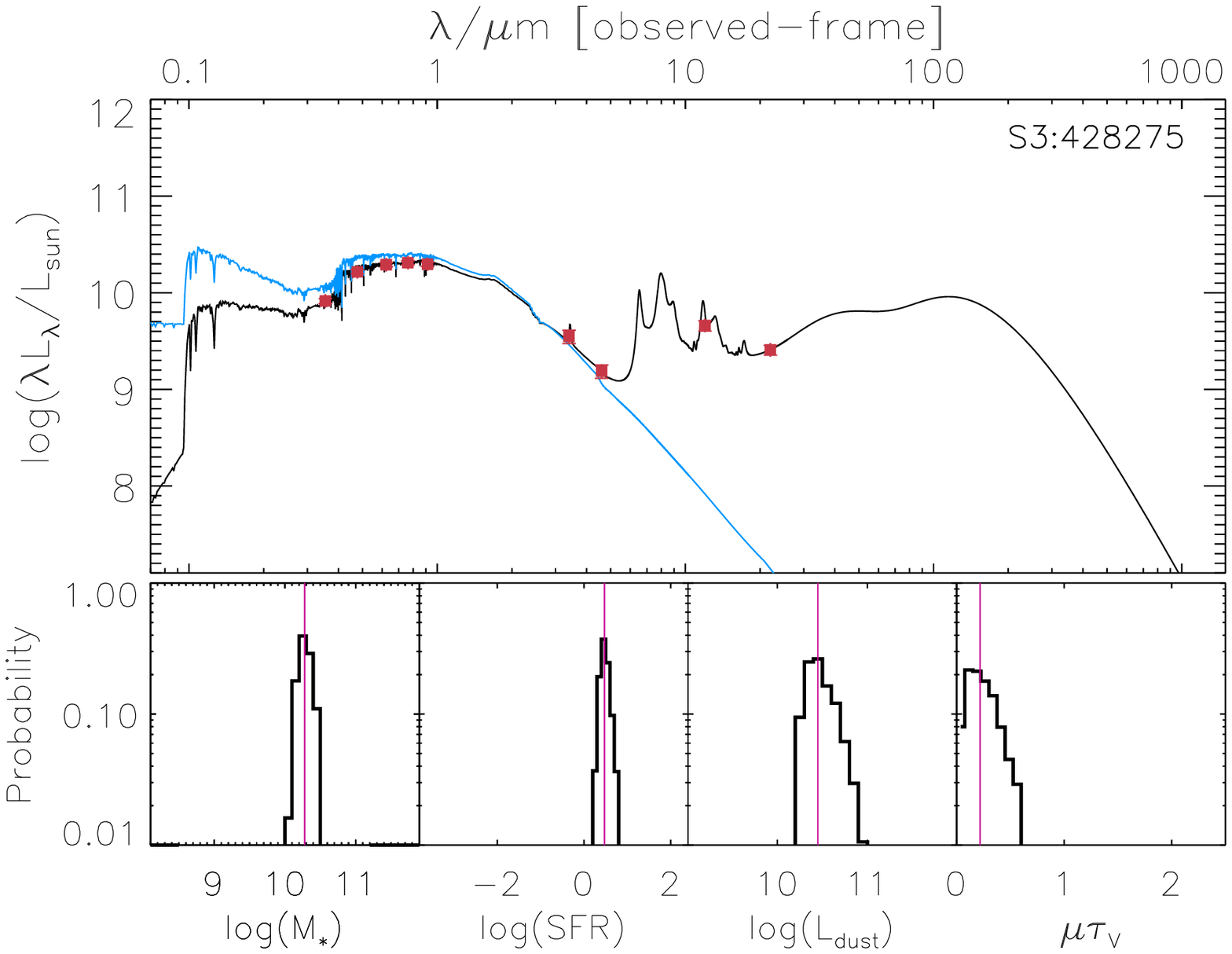}
\includegraphics[width=1\columnwidth]{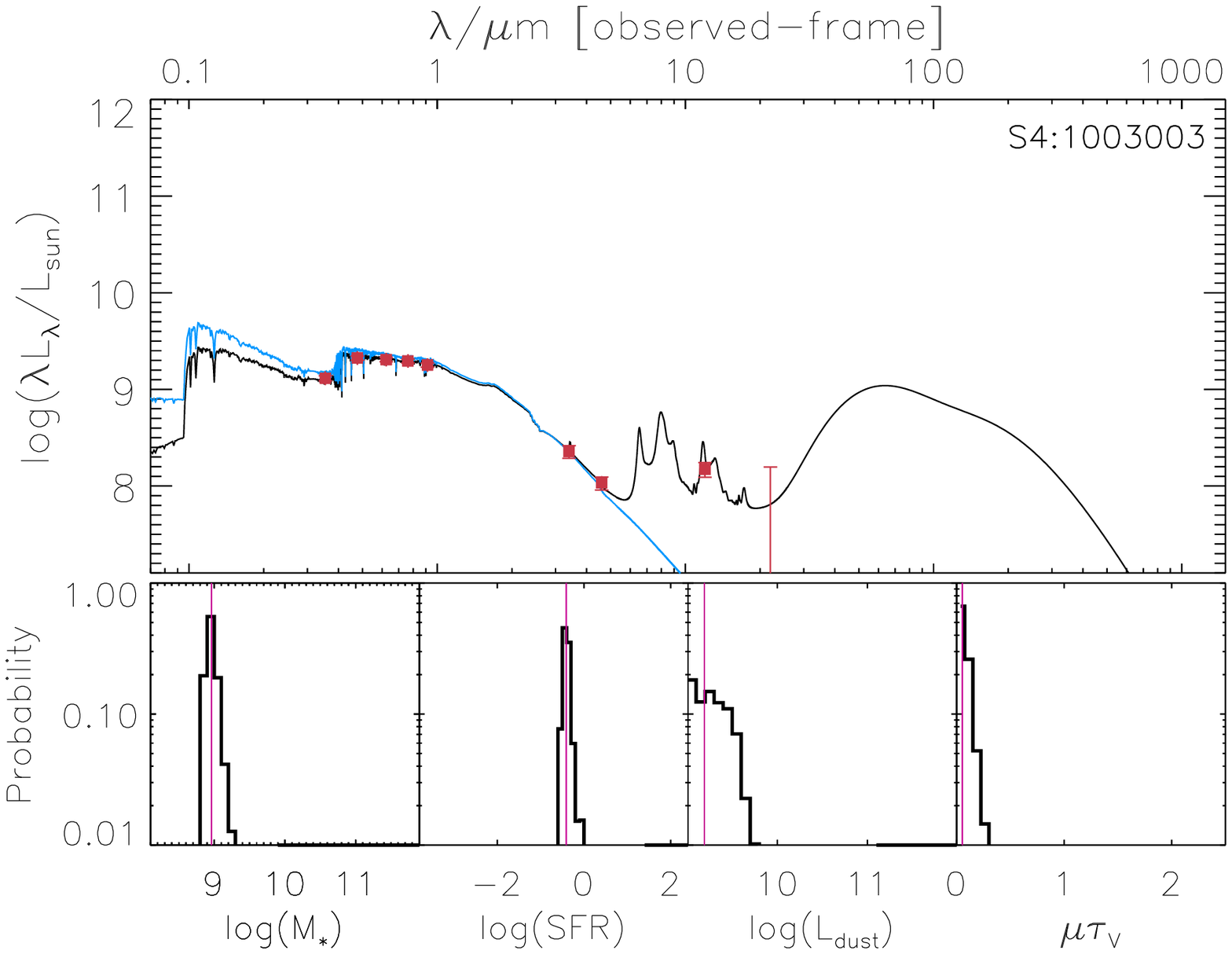}
\includegraphics[width=1\columnwidth]{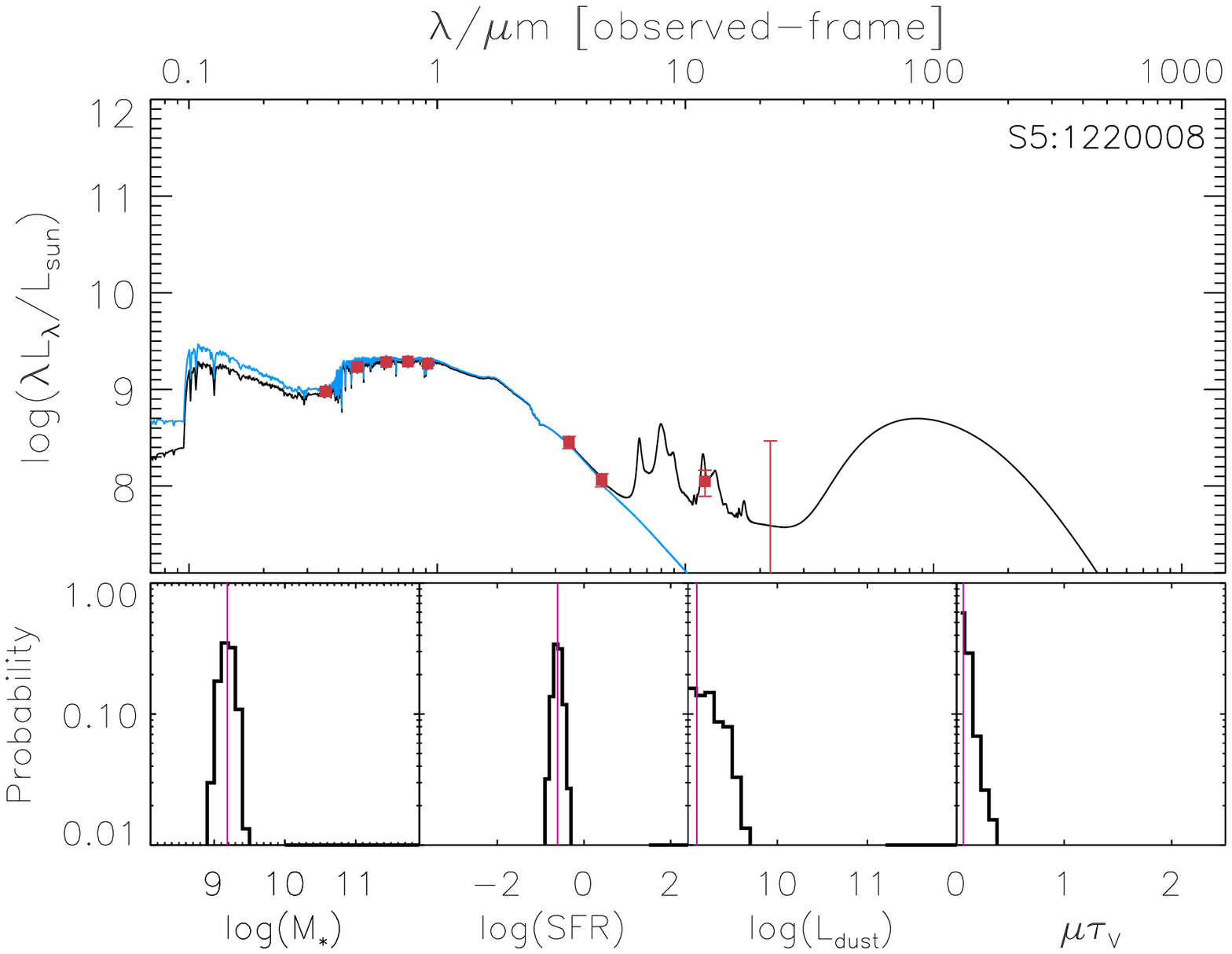}
\includegraphics[width=1\columnwidth]{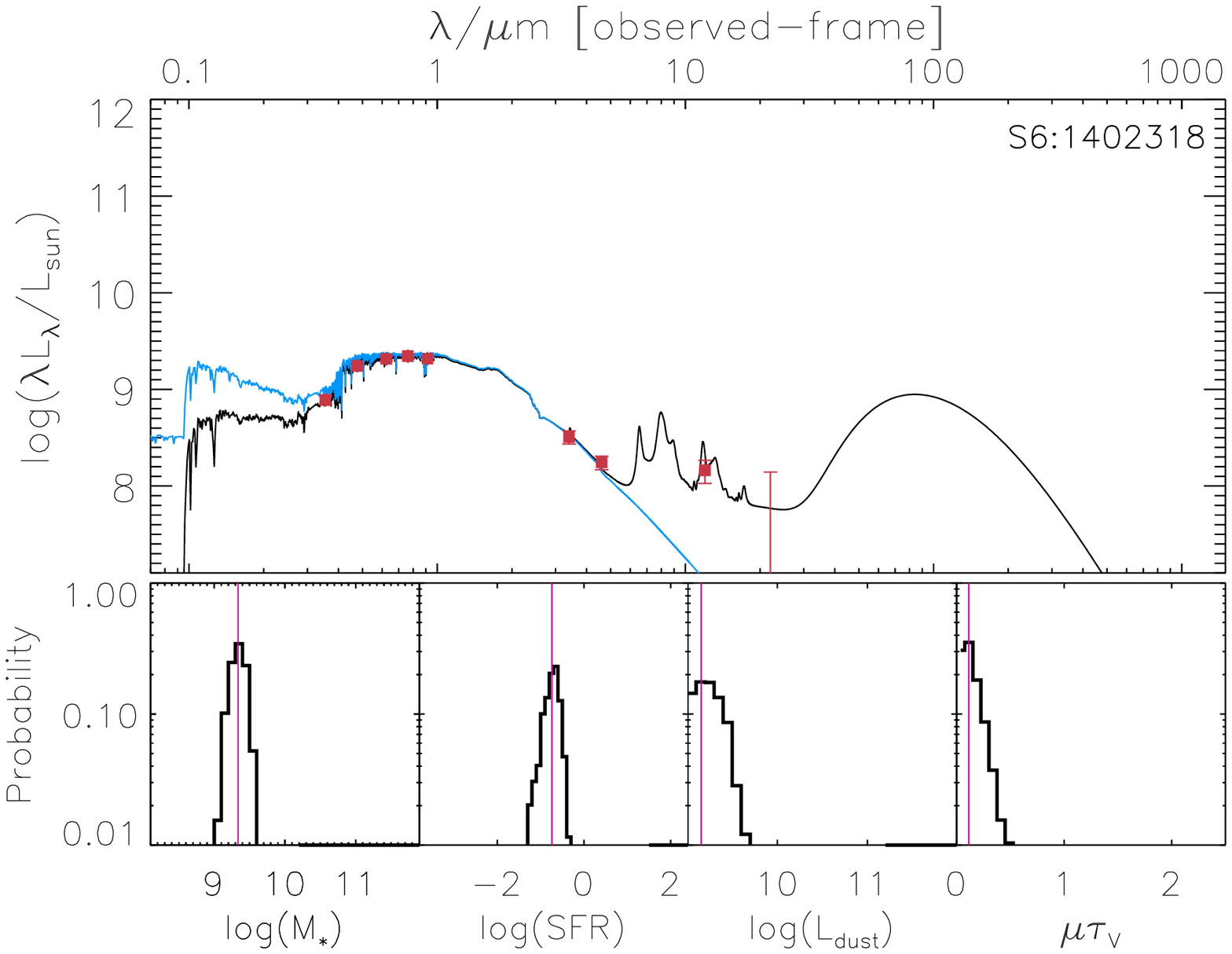}
\caption[]{Example SEDs and fitting results of six blue, star-forming
  galaxies , with only modest dust extinction in the optical.  See Figure \ref{massive} for an explanation of the panels, lines, and symbols.  The short-hand IDs (S1-6) correspond to
  the labels in Figure \ref{swfig_urz}.}
\label{blue}
\end{figure*}

\begin{figure*}
\centering
\includegraphics[width=1\columnwidth]{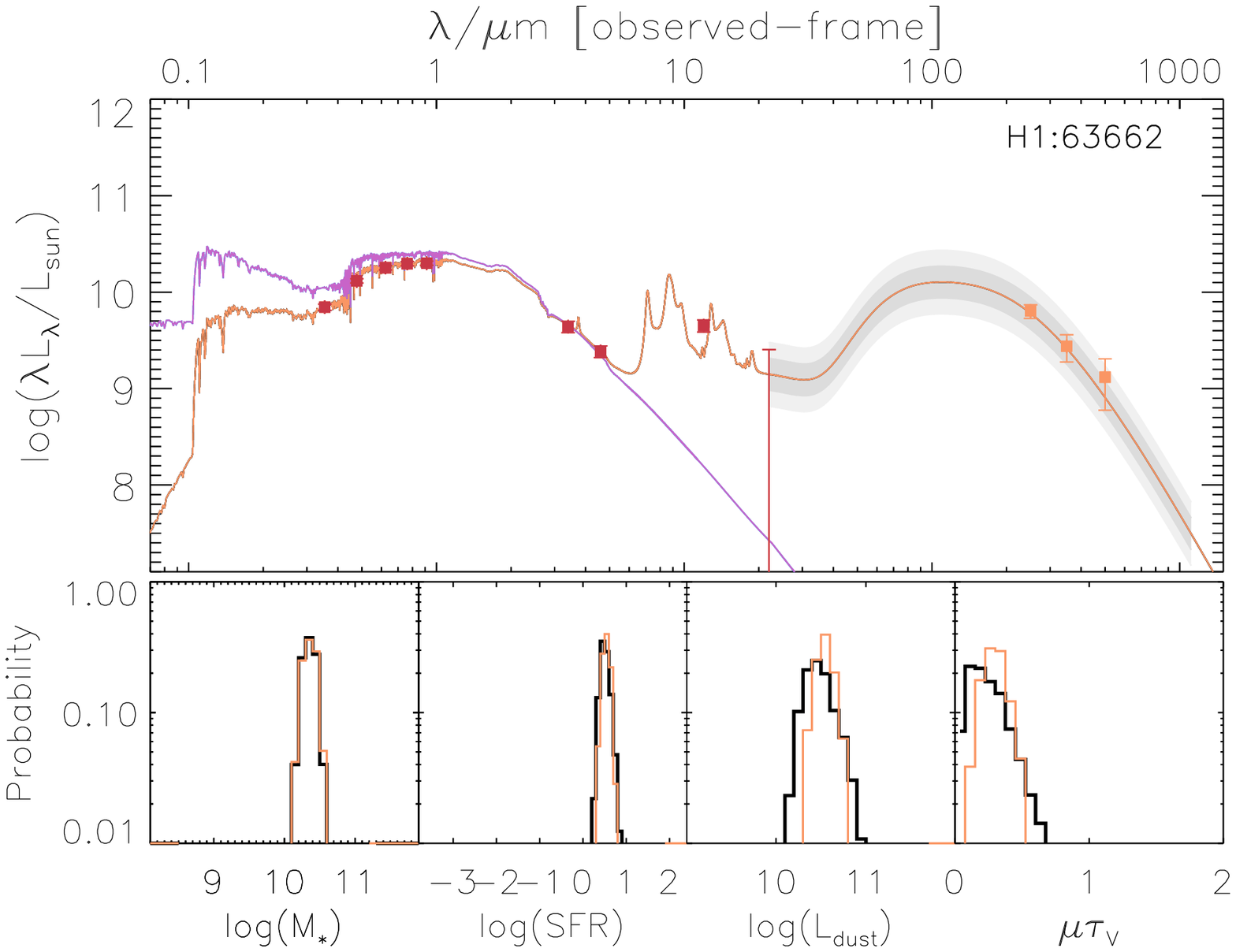}
\includegraphics[width=1\columnwidth]{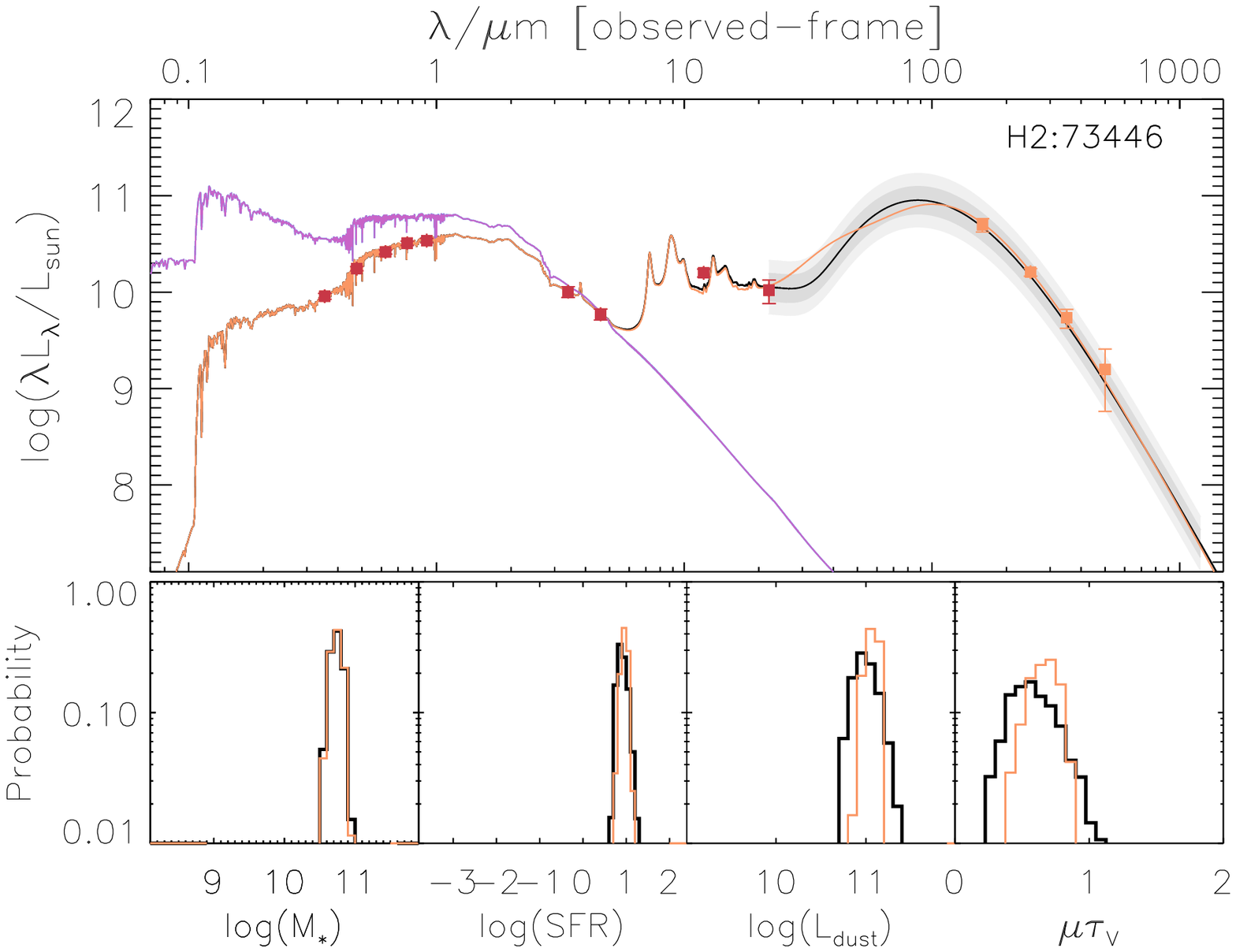}
\includegraphics[width=1\columnwidth]{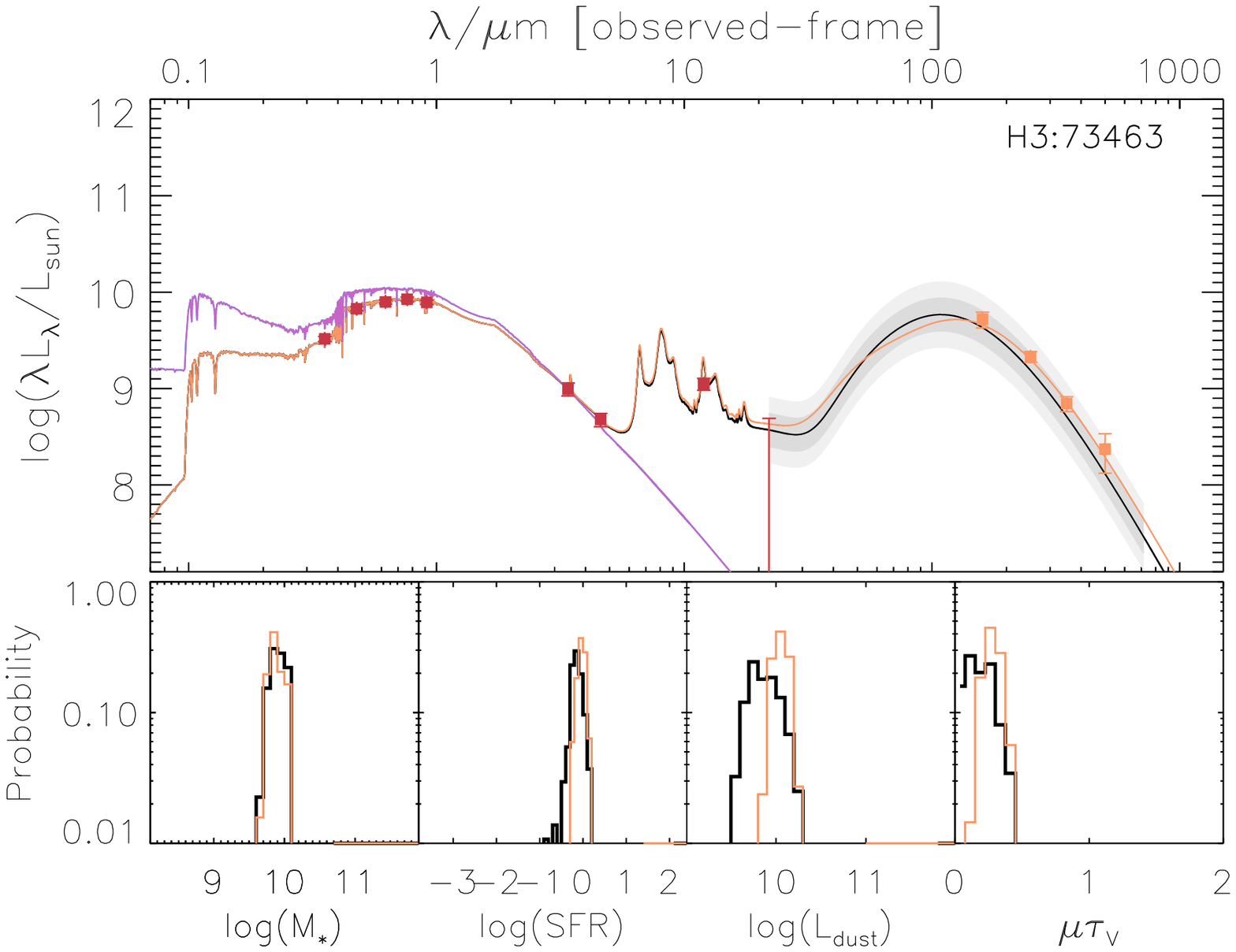}
\includegraphics[width=1\columnwidth]{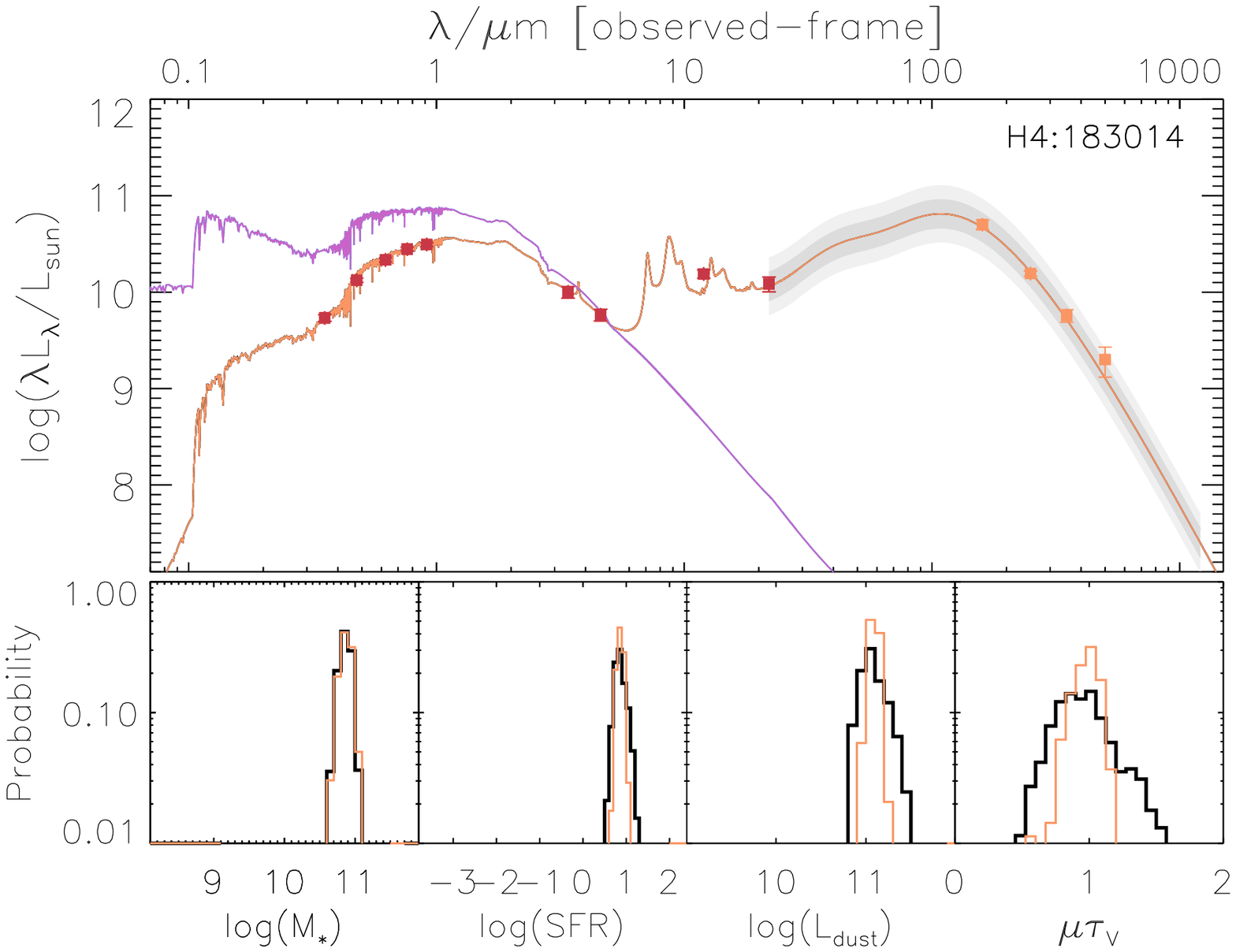}
\includegraphics[width=1\columnwidth]{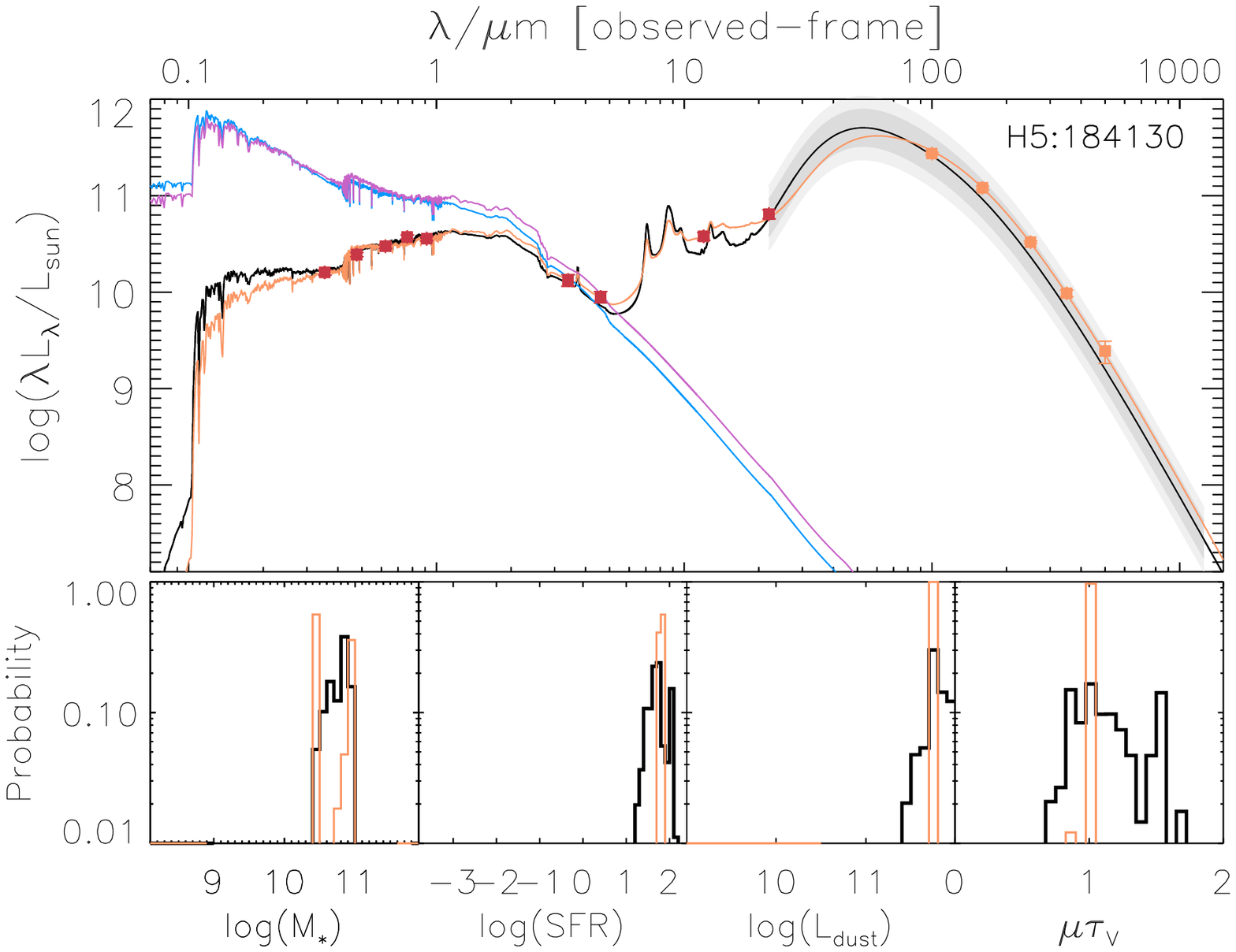}
\includegraphics[width=1\columnwidth]{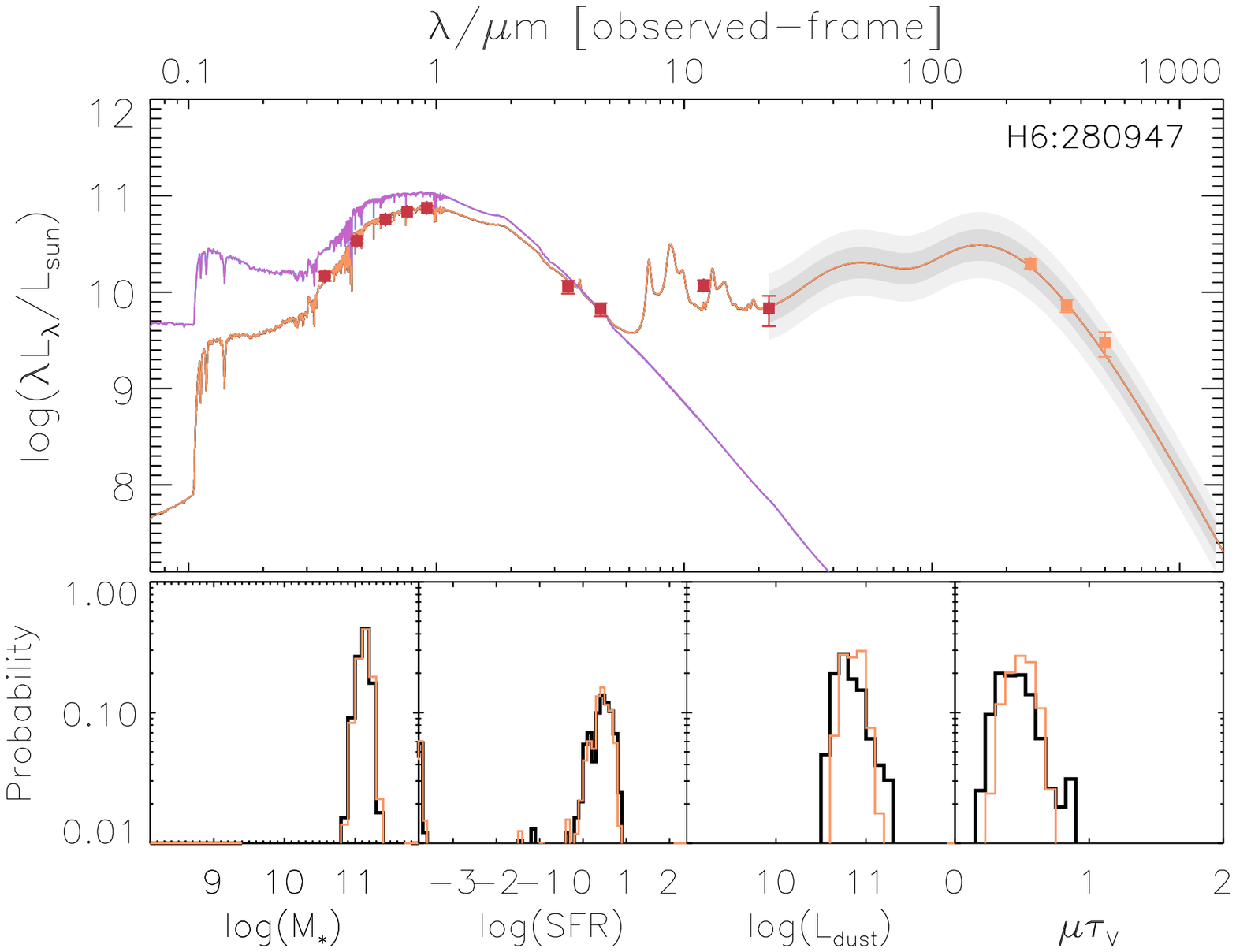}
\caption[]{Example SEDs and fitting results of six galaxies with
  Herschel-ATLAS photometry available (orange points) in addition 
  to the SDSS \& WISE (red points).  The blue and black lines
  represent fitting results to the SDSS + WISE data only (as in
  Figure \ref{massive}), and the dark (light) gray areas represent the
  68\% (95\%) confidence intervals for the far-IR dust SED {\it predicted} 
  from these fits. 
  The magenta and orange lines represent the best-fitting unattenuated and
  attenuated SEDs, respectively, when the Herschel-ATLAS SEDs are also 
  simultaneously fitted.  The smaller panels show the
  marginalized posterior probability distributions of four of the
  model parameters.   In all cases, the SDSS$+$WISE fitting results predict the
  Herschel photometry well and all results inferred without Herschel
  data are consistent with the results inferred with Herschel data.
  However, the constraints on the dust luminosity and dust attenuation parameter are
    tightened once Herschel photometry is included. The short-hand IDs
  (H1-6) correspond to the labels in Figure \ref{swfig_urz}.}
\label{swfig_hatlas}
\end{figure*}

\subsection{Sample SEDs}

The galaxies in our sample span a large range in stellar mass, star
formation rates and colors (Figure \ref{swfig_urz}).  In Figures
\ref{massive}, \ref{dusty}, and \ref{blue} we show examples of SEDs
and fitting results across the entire parameter space populated by our
sample.  Figure \ref{massive} shows six massive galaxies ($M_*\sim
10^{11}~M_{\odot}$ with little or no ongoing star formation ($\lesssim
1~M_{\odot}~\rm{yr}^{-1}$).  The characteristic red optical SEDs are
accompanied by non-neglible mid-IR emission arising from a combination
of dust heated by evolved stars and traces of star formation.  Figure
\ref{dusty} shows reddened galaxies with significant star formation
activity, where the IR luminosity often exceeds the optical/NIR
luminosity.  Figure \ref{blue} shows blue star-forming galaxies, with
little or no extinction.

\begin{figure*}[ht]
\centering
\includegraphics[width=1.0\textwidth]{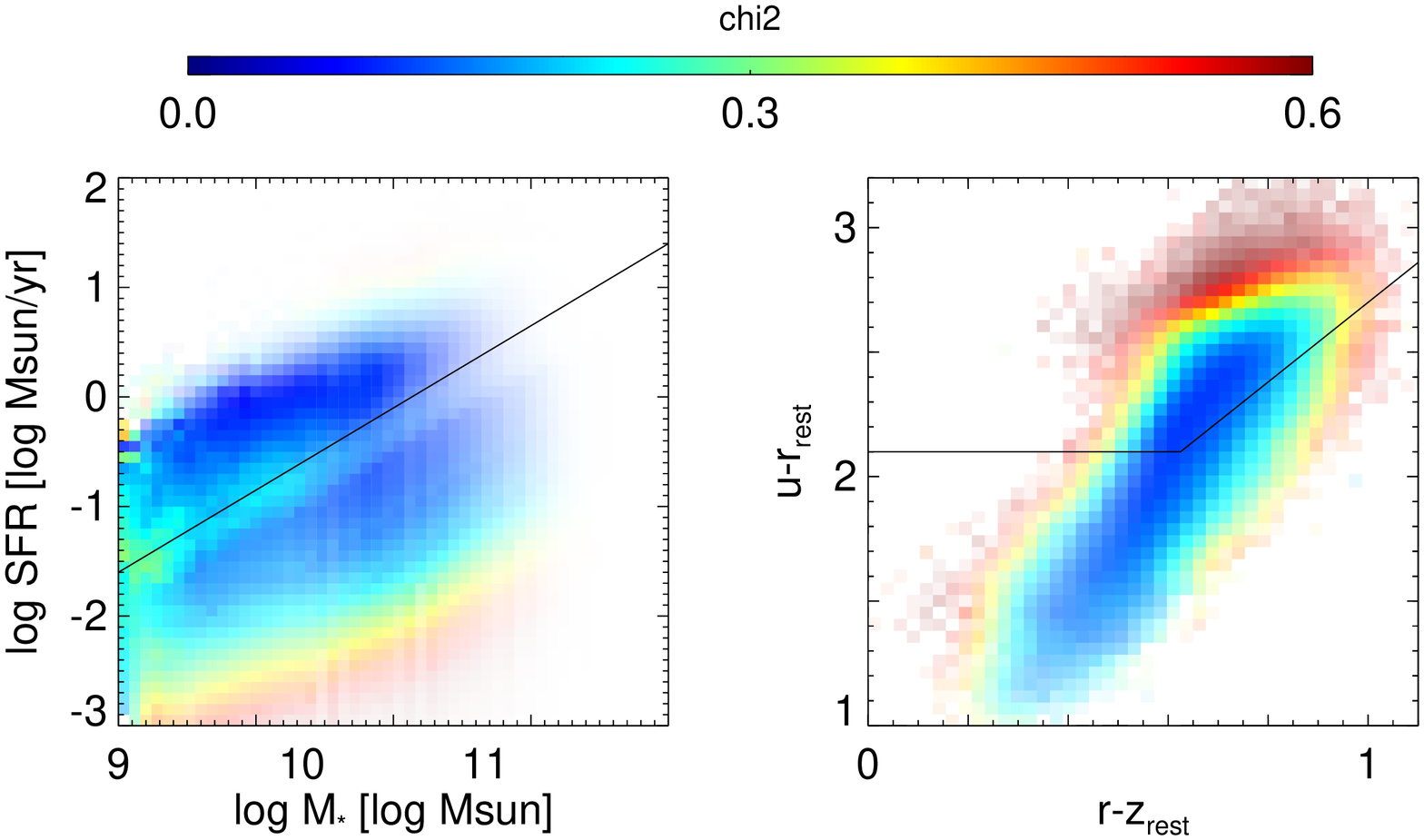} 
\caption[]{Quality of the SED fit, shown in dependence of SFR versus stellar mass (left; also see Figure \ref{swfig_mc_w} and Figure \ref{swfig_msfr}) and rest-frame $u-r$ vs. $r-z$ (right; also see Figure \ref{swfig_urz}). The color coding is the median chi-square value. Boxcar smoothing in two dimensions considering the uncertainties of the parameters is applied to the color coding and opacity. For star-forming galaxies, the distribution is about uniform.}
\label{swfig_ml}
\end{figure*}

To demonstrate the quality of the fits we show in Figure  \ref{swfig_ml} the chi-square distribution across parameter space ($M_*$, SFR, colors). $\chi^2$ is the the goodness-of fit parameter for the best-fitting model, but not, formally speaking, the reduced chi-squared of the best-fitting model as discussed by \citet{2012MNRAS.427..703S}. The generally low values of $\chi^2$ imply that our library of models is sufficiently extensive.

\subsection{Verification of total IR Luminosities with Herschel}

MAGPHYS balances the energy absorbed in the UV/optical and the energy
released in the IR, such that the inferred total IR luminosity can be
expected to be correct if the optical/NIR SED is accurate.  Still,
since the WISE 12$\micron$ and 22$\micron$ bands do not sample the
thermal peak of the dust emission, it is important to test the
robustness of the fitting results, in particular the star-formation
rates and dust luminosities.

\label{hatlas}
Herschel-ATLAS provides PACS and SPIRE photometry over the wavelength
range 100-500$\micron$. These observations sample the total dust
emission more directly than WISE.  For the subsample of 285 galaxies
that fall within the 16 square degree footprint of the publicly
available part of Herschel-ATLAS \citep{2010PASP..122..499E,
  2010MNRAS.409...38I, 2011MNRAS.415..911P, 2011MNRAS.415.2336R,
  2011MNRAS.416..857S} we verify the precision and accuracy of our
inferred total IR luminosity.

\begin{figure}
\centering
\includegraphics[width=1.0\columnwidth]{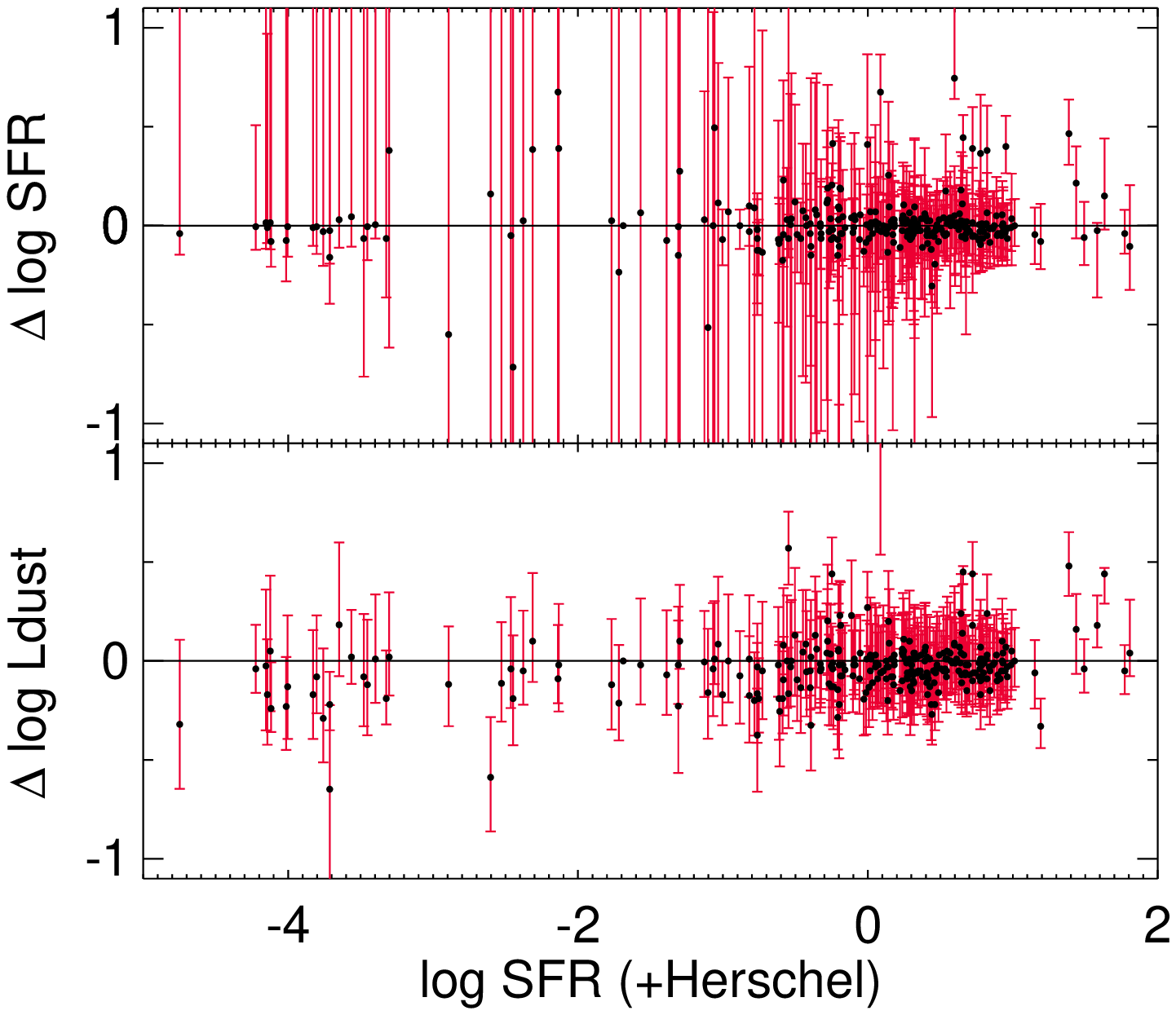} 
\caption[]{{\it Top:} SFR from the SED fits with only optical (SDSS)
  and near/mid-IR (WISE) data minus SFR from fits that also include
  far-IR Herschel data vs. SFR from fits with Herschel data.  The
  objects included here are those that overlap between our
  SDSS-selected sample and Herschel-ATLAS.  
  Red error bars indicate 68\% confidence intervals.
  {\it Bottom:} Same as above, but for dust luminosity.  Our SDSS+WISE
  based fitting results are generally consistent with the
  SDSS+WISE+Herschel-based fitting results.}
\label{swfig_hatlas_compare}
\end{figure}

Our SDSS+WISE-based models succesfully predict of total IR
luminosities and star-formation rates from SDSS+WISE+Herschel
photometry, as illustrated in Figure \ref{swfig_hatlas}).  Even though
the marginalized probability distributions for the dust luminosity and
star-formation rate are tightened when Herschel photometry is added,
we find no systematic offset (0.00 dex) between the star-formation
rates.  Furthermore, the scatter (0.19 dex) is consistent with the
formal confidence intervals (see Figure~\ref{swfig_hatlas_compare}).
The uncertainties of $SFR\sim10^{-4} M_{\odot}/yr$ galaxies are large and dust luminosities are poorly constrained.

\subsection{The Public Catalog}

We provide two public
catalogs{\footnote{\url{http://irfu.cea.fr/pisp/yu-yen.chang/sw.html}}}.
They both contain all 858,365 galaxies from the SDSS spectroscopic
galaxy sample with good redshift measurements as described in
Section~\ref{sample}.  Table~\ref{tab_cat_in} contains the input data
(ID, redshift, fluxes, Galactic extinction).  Table~\ref{tab_cat_out}
contains the modeling results from MAGPHYS, $V_{max}$ (see below),
rest-frame luminosities, and a set of flags.  
We recommend to use the MAGPHYS modeling results for objects with
FLAG=1 (633,205 out of 858,365).  These are all $z<0.2$ galaxies with
reliable aperture corrections based on size measurements from
\citet{2011ApJS..196...11S} ($FLAG\_R=1$), good WISE photometry
($FLAG\_W?=1$ or $2$), and good-quality SED fits ($FLAG\_CHI2=1$).
This primary sample contains 91.40\% of all SDSS galaxies at $z<0.2$.

Our $V_{max}$ calculation only includes the maximum redshift at which galaxies would be retained in the sample.  No low-redshift limit, related to the SDSS bright-magnitude limit, is taken into account.  For our purposes of illustrating the distribution of SFR and stellar mass this does not matter given the large volume, but for individual, low-redshift, high-luminosity galaxies the user should keep in mind that our $V_{max}$ cannot be directly used.
\label{catalog}


\begin{figure*}[h]
\centering
\includegraphics[width=2.0\columnwidth]{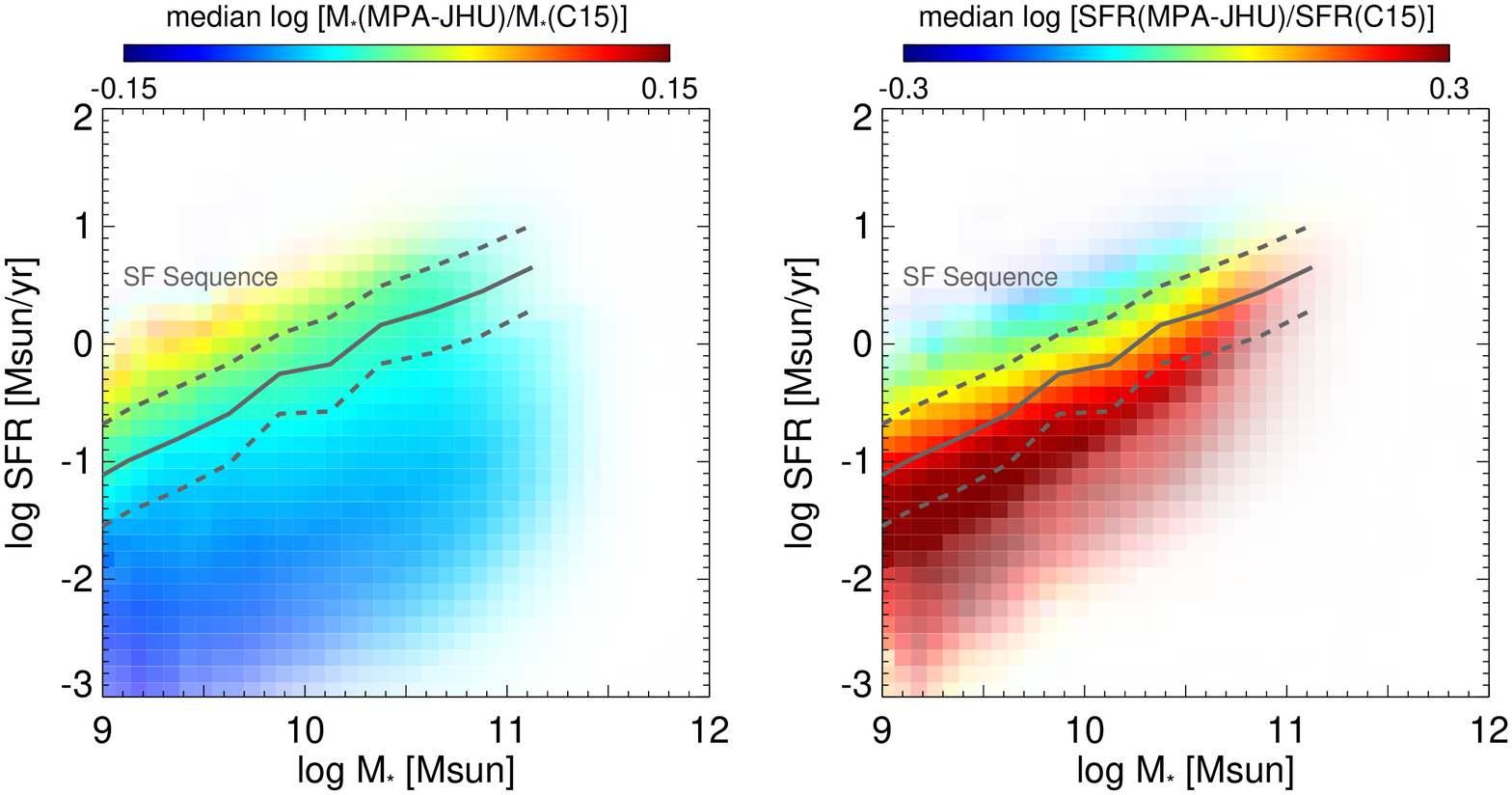}
\includegraphics[width=2.0\columnwidth]{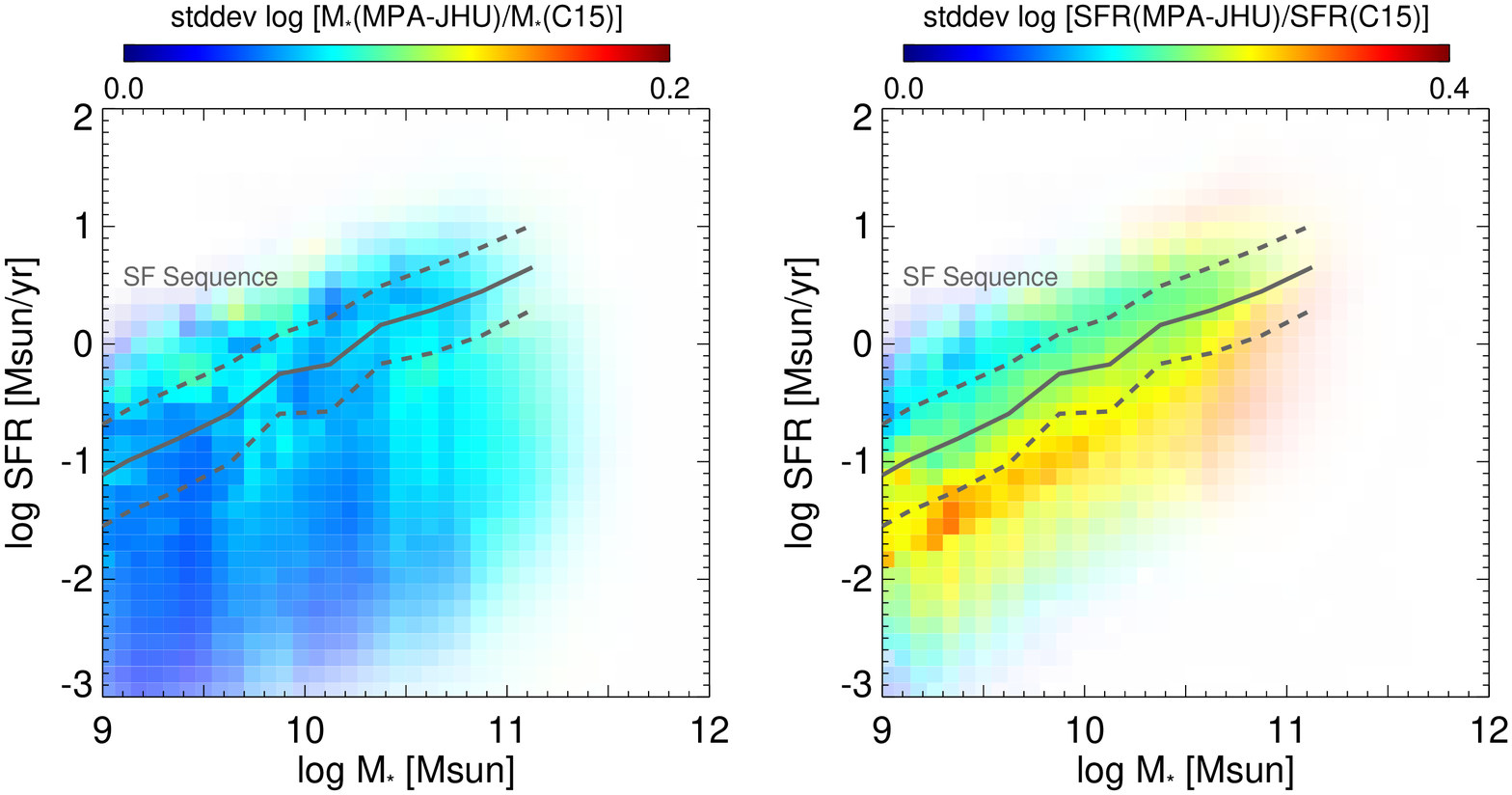}
\caption[]{Comparison between the MPA-JHU values for $M_*$ and
  SFR and the values derived here (C15).  
  {\it Top left:} median difference between the stellar mass
  estimates.  These are always small ($\lesssim~0.10$~dex).  {\it
    Bottom left:} scatter in the stellar mass estimates; generally
  $\lesssim0.15$~dex. {\it Top right:} median difference in SFR. Here
  we only show star-forming galaxies selected the color-color
  classification given in the right-hand panel of Figure
  \ref{swfig_urz}.  Across the sample the offset is 0.22 dex. It is
  strongly SFR-dependent, and ranges from -0.3 dex to +0.3 dex.  {\it
    Bottom right:} scatter in the SFR estimates, typically 0.3 dex or a factor of two.
   The grey lines shows the star-forming sequence in Figure~\ref{swfig_msfr}.}
\label{swfig_mc_w}
\end{figure*}

\section{Discussion}
\label{sec4}

\subsection{Comparison with Brinchmann et al. (2004)} 
\label{c_b04}
Stellar masses and SFRs from the MPA-JHU catalogs
\citep{2004MNRAS.351.1151B} are widely used.  The differences between
their and our stellar mass estimates are the addition of WISE
$3.4\micron$ and $4.5\micron$ photometry as tracers of stellar mass,
and the updated Galactic extinction correction from
\citet{2011ApJ...737..103S}.  Furthermore, the simultaneous fitting of
mid-IR photometry provides a fundamentally different constraint on the
dust properties of galaxies than optical photometry and emission line
ratios. 
In addition, dust attenuation laws are also different. In \citet{2004MNRAS.351.1151B} the adopted slope for the attenuation curve is -0.7, following \citet{2000ApJ...539..718C}. MAGPHYS has the same slope for diffuse dust, but a steeper slope (-1.3) for birth clouds.
Despite these differences, offsets between the two sets of
mass estimates are small and insignificant (left-hand panel,
Figure~\ref{swfig_mc_w}).  The reasons for this are that the same
stellar population models and star-formation histories are used and
that the impact of the WISE photometry is limited due to the
relatively large uncertainties on the total flux measurements.

Larger differences are seen for the SFR estimates.  As opposed to the
stellar mass estimates, the SFR estimates rely on wholly different
tracers in the two cases.  Selecting star-forming galaxies by their
location in the color-color diagram (Figure \ref{swfig_urz}), we find
a median offset across the sample of 0.22 dex and a scatter of
$\sim0.3$~dex.  This offset does not depend on mass or redshift, or
even on whether the galaxies have significant 12$\mu$m and 22$\mu$m
detections, but strongly varies with SFR (Figure \ref{swfig_mc_w}):
for high-SFR galaxies our values are large compared to
\citet{2004MNRAS.351.1151B}, for low-SFR galaxies our values are
small. The anti-correlation between the changes in $M_*$ and SFR is the result of the underlying anti-correlation between mass to light ratio and sSFR.

\subsection{12$\mu$m and 22$\mu$m luminosities as SFR indicators}

Another consequence is that our 12$\mu$m
and 22$\mu$m SFR conversions are lower by, respectively, 0.22 dex and 0.10 dex
compared to the conversion based on a comparison with the
\citet{2004MNRAS.351.1151B} estimates, as carried out by
\citet{2013ApJ...774...62L}.

\begin{figure*}[]
\centering
\includegraphics[width=1.0\textwidth]{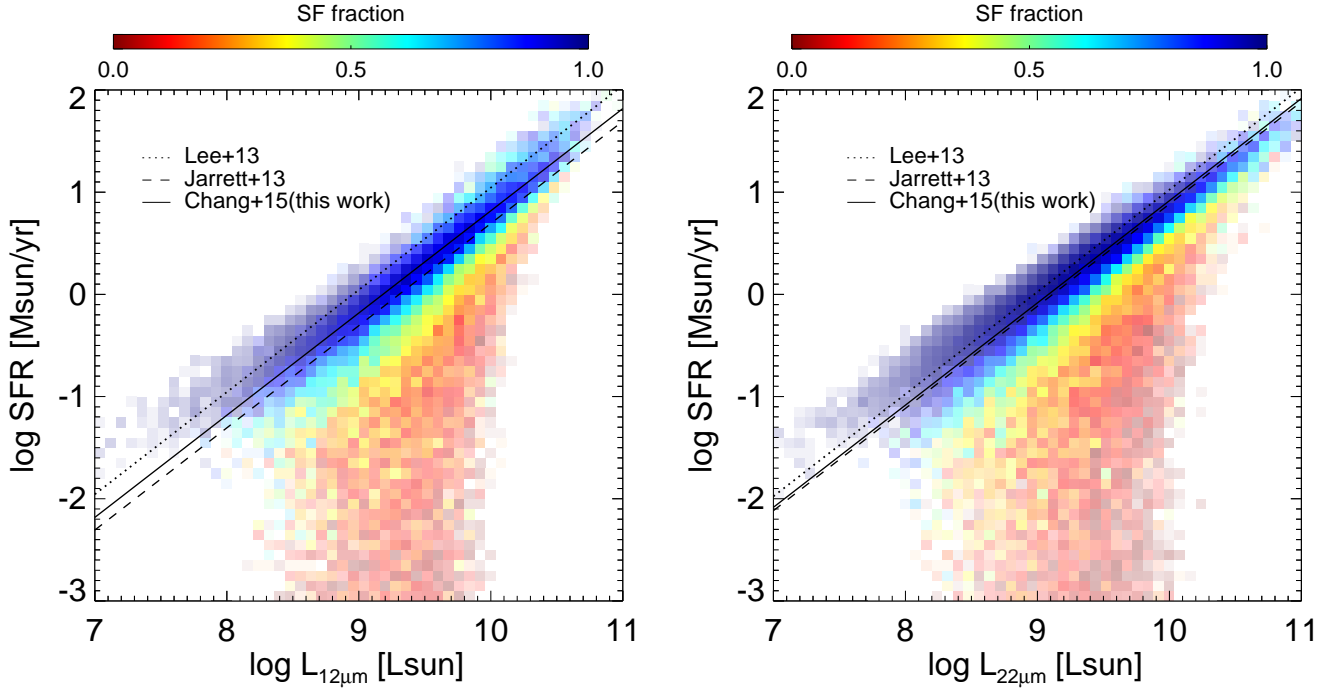} 
\caption[]{ SFR vs.~rest-frame 12$\mu$m and 22$\mu$m luminosities for
  galaxies that have significant detections in both WISE bands.  The data
  are binned and color-coded according to the fraction of star-forming
  galaxies as defined in the color-color diagram shown in Figure
  \ref{swfig_urz}.  For star-forming galaxies we produce a linear
  fit, as i ndicated by the solid lines.  For comparison we show
  equivalent fits from \citet{2013ApJ...774...62L} and \citet{2013AJ....145....6J}. 
  The SFRs are matched to \citet{2003PASP..115..763C} IMF.}
\label{swfig_1222sfr}
\end{figure*}

\label{l12_l22_SFR}
In Figure \ref{swfig_1222sfr} we show the correlations between
12$\mu$m and 22$\mu$m luminosities and our SFR estimates.  The large
downward scatter is due to mid-IR radiation that is not associated with
star formation.  In particular, the mid-IR luminosities of quiescent
galaxies do not reflect star formation activity, but rather
circumstellar dust and PAHs heated by evolved stars
\citep[e.g.,][]{2006ApJ...639L..55B}.  Even for many galaxies that are
star-forming according to our definition -- separated in color-color
space (Figure \ref{swfig_urz}; SF galaxies: $(u-r)_{rest} < 2.1$ or $(u-r)_{rest} < 1.6 \times (r-z)_{rest}+1.1$) -- a large fraction of the mid-IR
luminosity is not attributed to star formation.  Keeping this in mind,
we derive the following conversions from mid-IR luminosity to SFR:

\begin{equation}
\log{SFR/(M_{\odot}~yr^{-1})} = \log{L_{12}/L_{\odot}} - 9.18
\end{equation} 

and

\begin{equation}
\log{SFR/(M_{\odot}~yr^{-1})} = \log{L_{22}/L_{\odot}} - 9.08,
\end{equation} 

These relationships are determined by calculating the median values of
the SFR in bins of 0.25 dex wide bins in luminosity and performing a
linear fit to these median values, weighing by the inverse of the
square root of the numer of objects in the bins.  The upward scatter
(84\%-ile minus median) in SFR is 0.20 dex, while the downward scatter
(median minus 16\%-ile) is 0.60 dex, reflecting non-SF contributions.
The scatter obviously depends on luminosity: above $10^{10}~L_{\odot}$
the downward scatter is also 0.30 dex.

The 12$\mu$m and 22$\mu$m values are 0.13 and 0.04 dex higher than the conversions provided by  \citet{2013AJ....145....6J}, who based their SFRs on GALEX UV fluxes and the standard Spitzer/MIPS 24$\mu$m calibration from \citet{2009ApJ...692..556R}. The relatively good agreement with the \citet{2013AJ....145....6J} calibrations is encouraging, as their sample consists of 17 well-resolved, nearby galaxies for which measurement uncertainties are minimal. (In fact, the small difference is consistent with the random variation expected in the average for small sample of 17 objects.) Our calibration generalizes their result by extending the dynamic range in stellar mass and star-formation rate by an order of magnitude upward and using a sample of several hundred thousand galaxies.

Our 12$\mu$m and 22$\mu$m SFR conversions are lower by, respectively, 0.22 dex and 0.10 dex compared to \citet{2013ApJ...774...62L}, who use the \citet{2004MNRAS.351.1151B} SFRs.\footnote{These differences include the appropriate correction from Salpeter to Chabrier IMF for the \citet{2013ApJ...774...62L} conversions.} Part of this offset can be attributed to the systematic differences between our SFR estimates and those of \citet{2004MNRAS.351.1151B} (see Sec \ref{c_b04}). Additional factors are the differences in sample selection and modeling technique. We select star-forming galaxies on the basis of their optical colors, whereas \citet{2013ApJ...774...62L} select galaxies with large 22$\mu$m luminosities. Moreover, not all mid-IR radiation traces SF, which is taken into account in our modeling, whereas \citet{2013ApJ...774...62L} simply use total mid-IR luminosity as a SF tracer.

\subsection{A Fresh View on Bimodality}

\begin{figure}[ht]
\centering
\includegraphics[width=1.00\columnwidth]{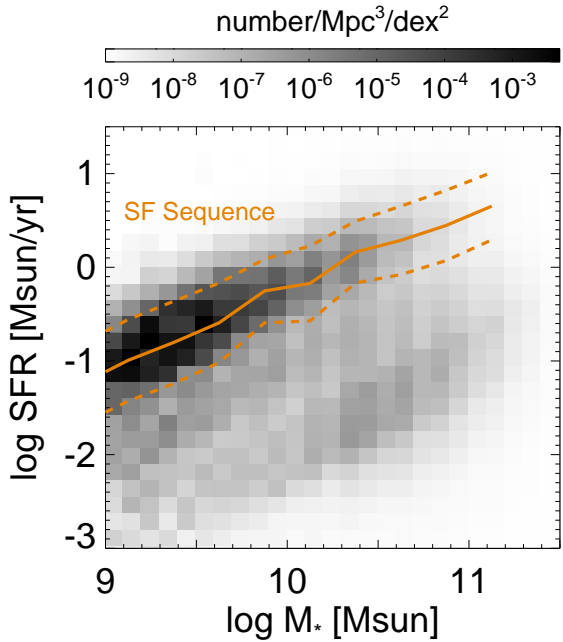} 
\caption{Distribution of galaxies above our (redshift-dependent) mass
  completness limit in the SFR vs $M_*$ plane.  The mass limit
  described in Section~\ref{msfr} is such that all
  galaxies on or near the the SF-ing sequence have significant
  detections at 12$\mu$m.  The star-forming sequence is shown in
  orange, where the dashed lines show the
  1$\sigma$~scatter, as explained in Section~\ref{msfr}.
  In order to visualize the SFR bimodality we adopt the 1-$\sigma$
  upper limits on the SFR for galaxies where the SFR estimate is more
  uncertain than 0.39 dex.  Consequently, quiescent galaxies in this
  plot are placed at the upper limit of the true SFR.  The gray scale
  represents the number of galaxies after weighing by $1/V_{max}$.  }
\label{swfig_msfr}
\end{figure}

\label{bimodality}
Assessing bi-modality in the galaxy population requires a sample that
is corrected for completeness and sufficient sensitivity in the
star-formation tracer to separate star-forming and passive galaxies.
We compare the magnitude in $r$ band with the WISE bands at different redshift and find that the mass completeness limit of our sample is set by the SDSS $r$-band spectroscopic limit for all redshifts: the fraction of galaxies above the mass limit without WISE counterparts is less than 1\%.
The completeness limit at a given redshift is
found by identifying the most massive galaxies with magnitudes near
the limit.  As such we find that the mass completeness limit depends
on redshift as $\log M_{limit} = 10.6 + 2.28 \log (z / 0.1)$.  The
mass limit is such that actively SF-ing galaxies always have
significant 12$\mu$m detections; 
that is, if a galaxy is not detected at 12 $\mu$m it must be a quiescent galaxy.  In other words, the upper limit on SFR is useful is the sense that it allows us to distinguish between SF-ing and quiescent galaxies.  
Using the mass limit, a
galaxy of a given mass is assigned a $V_{max}$ value assuming a survey
area of 8,032 square degrees.  These values are given in Table~2.

\label{msfr}
In Figure \ref{swfig_msfr} we show the SFR-$M_*$ distribution of the
363,774 galaxies above our mass limit and with FLAG = 1 (see Section
\ref{catalog}).  The sequence of star-forming galaxies (a.k.a.~the
Main Sequence) -- indicated in orange -- is quantified by the median
in 0.2-wide stellar mass bins.  We fit these median values with a
power law and find:
\begin{equation}
  \log{SFR/(M_{\odot} yr^{-1})} = 0.80\log{M_*/(10^{10}~M_{\odot})} - 0.23.
\end{equation} 

We define the scatter in the Main Sequence as the 84\%-50\%-ile range,
which is 0.39 dex.  The downward scatter (50\%-16\%-ile range) is
larger (0.64 dex), but this number is difficult to interpret due to
the imperfect separation of star-forming and quiescent galaxies.  This
uncertainty is not (only) due to limited fidelity in our color-color
classification, but also due to the natural variation in SF activity.

For comparison, when we apply our fitting technique
using the \citet{2004MNRAS.351.1151B} values for stellar masses and SFRs, we find:
\begin{equation}
  \log{SFR/(M_{\odot} yr^{-1})} = 0.75\log{M_*/(10^{10}~M_{\odot})} - 0.03.
\end{equation} 
Our fit is slightly steeper, but more importantly, has a 0.2 dex
higher intercept at $M_* = 10^{10}~M_{\odot}$ and a larger scatter
(0.39 dex vs.~0.31 dex).  The larger scatter can perhaps be attributed
to the more direct measurement of highly obscured star formation when
using mid-IR luminosities as a tracer as opposed to the optical
emission lines.
We note that the random uncertainties in our SFR estimates (typically, 
0.18 dex) are much smaller than the scatter, indicating a large intrinsic
scatter around the ``main sequence'' relation.  This is in stark contrast
with the measurements from \citet{2004MNRAS.351.1151B}: their typical uncertainty 
is 0.30 dex -- the same as the scatter -- implying zero intrinsic scatter.  
We conclude that the the star-forming sequence as revealed by the WISE photometry 
is less tight than what is inferred from optical SDSS spectra.

The bi-modality seen in Figure \ref{swfig_msfr} in our inferred
SFR-$M_*$ distribution can be directly compared with Figure 17 and 24
from \citet{2004MNRAS.351.1151B}, where the SFRs are inferred from
nebular emission lines strengths, and with Figure 15 from
\citet{2007ApJS..173..267S}, where the SFRs are inferred from UV
luminosities.  The color coding used in Figure \ref{swfig_urz} is
based on the separation of star forming and non-star forming galaxies
in Figure \ref{swfig_msfr}: non-star forming galaxies are those that
lie below the sequence by 2 times the scatter or more.  This confirms
that the two-color diagram is a simple yet very effective tool to
separate dusty from passive galaxies.

\begin{figure*}[ht]
\centering
\includegraphics[width=1.0\textwidth]{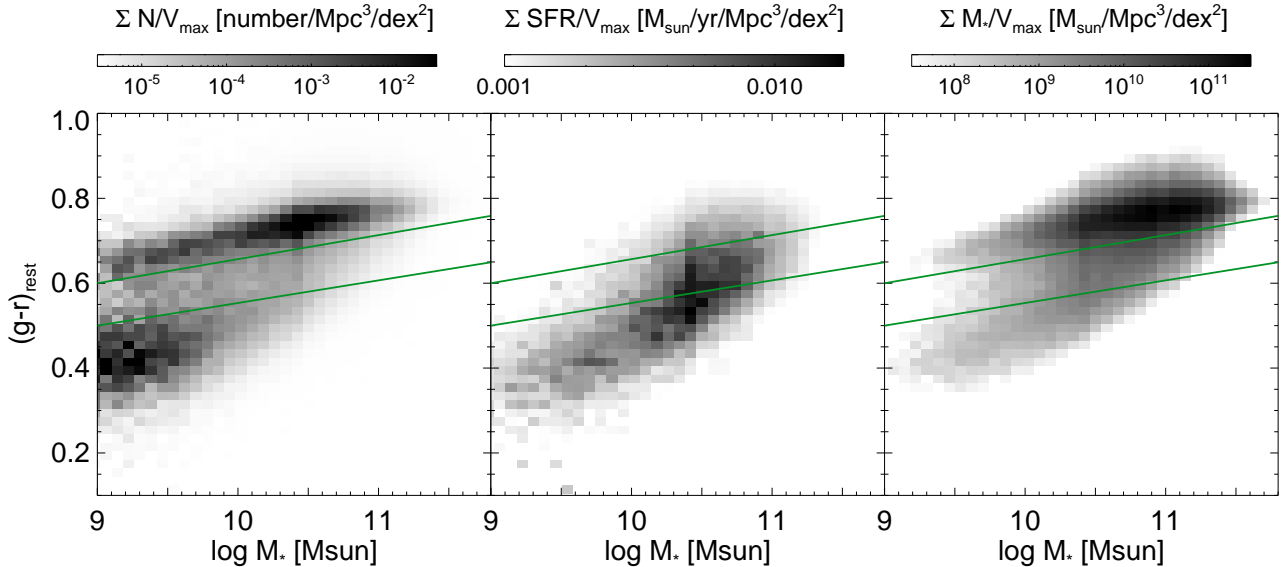} 
\caption[]{Left: comoving number density distribution of galaxies in the
  stellar mass vs. rest-frame $g-r$ plane. The grey scale
  reflects the numbers of galaxies in each bin after weighing
  with $1/V_{max}$. Middle: the grey scale reflects the SFR density. 
  Right: the grey scale reflects the stellar mass density.  
  Apparently, the green valley provides the most fertile ground for 
  star formation as indicated by the green line.}
\label{swfig_mcolor}
\end{figure*}

Finally, it is of interest to explore the distribution of SF in
relation to traditional tools to probe SF activity across the galaxy
population, such as the color-mass diagram, the BPT diagram \citep{1981PASP...93....5B}, 
and the H$\delta$-D4000 relation \citep{2003MNRAS.341...33K}.  In
Figure \ref{swfig_mcolor} we show the $g-r$ vs.~stellar mass distribution in three
different ways: weighed by number density, weighed by SFR density, and
weighed by stellar mass density.  The first panel shows the well-known
bimodality in the form of a red sequence and a blue cloud. We have
also indicated the intermediate region often called the green valley,
which has been argued to contain galaxies that are transitioning from
blue to red \citep[e.g., ][]{2007ApJS..173..512S}.  However, as most recently shown
by \citet{2015MNRAS.446.2144T}, many galaxies in this region are reddened
because of dust \citep[also see, e.g., ][]{2006MNRAS.373.1389C} and not
primarily due to a reduced level of SF.  This is illustrated in the
second panel, which shows that the galaxies in the green valley region
-- and not the more numerous, fainter blue galaxies -- dominate the
total SF budget of the present-day universe.  On the other hand, these
massive, SF-ing galaxies do not dominate the stellar mass budget, as
is shown in the third panel.  The absence of a dominant population of
massive, SF-ing galaxies and a general dearth of massive, disk-like
galaxies \citep{2009ApJ...706L.120V} implies that their SFRs
cannot be sustained for long \citep{2014MNRAS.440..889S}.

\begin{figure*}[ht]
\centering
\includegraphics[width=0.9\textwidth]{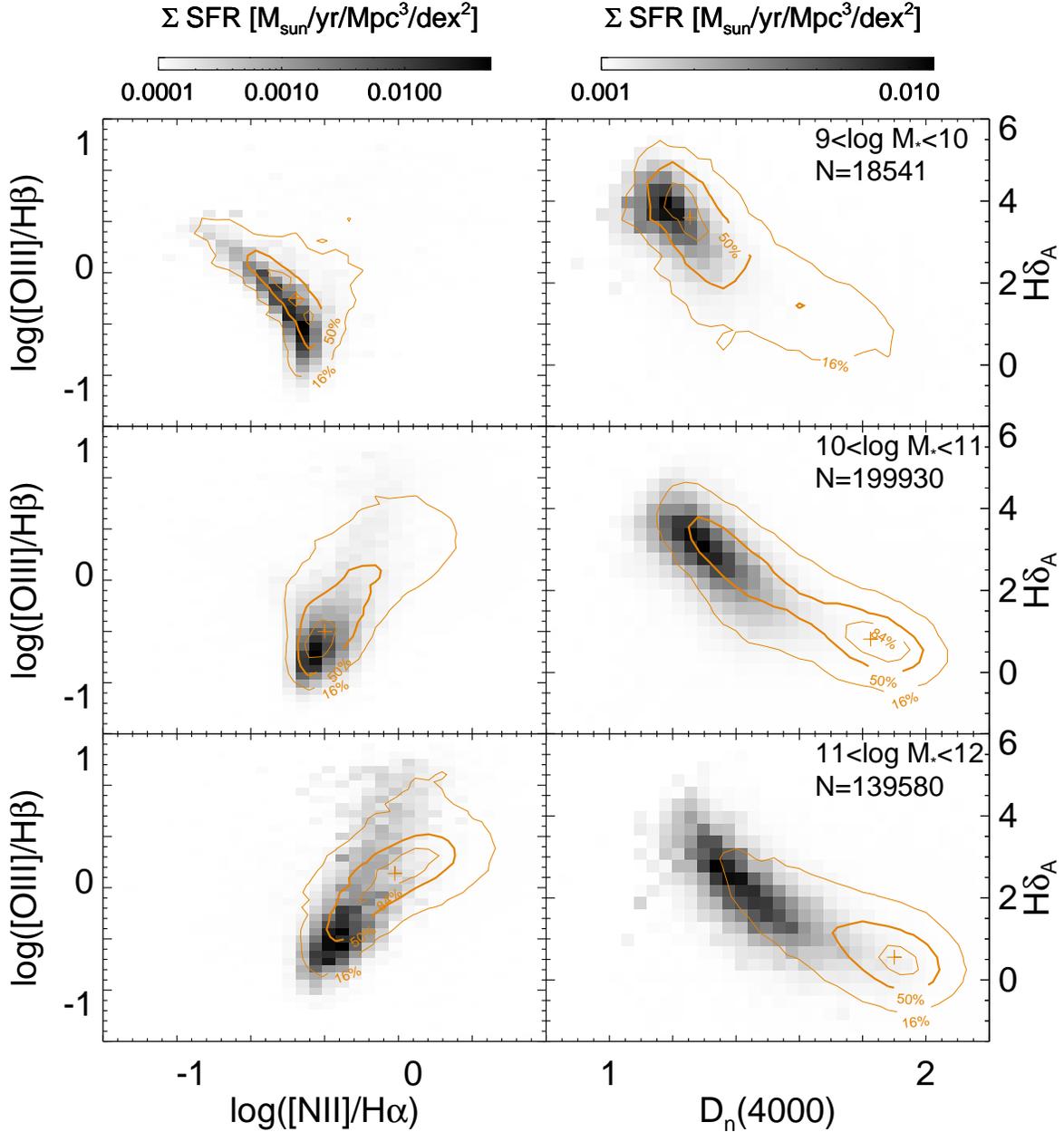} 
\caption[]{Number density (contours) and SFR density distribution (grey scale) in the BPT diagram (left, only galaxies with detected emission lines) and $ H\delta$-$D_n(4000)$ (right).  From top to bottom we show 3 different mass bins. }
\label{swfig_d4000hda}
\end{figure*}

In Figure \ref{swfig_d4000hda} we show the SFR distribution in the BPT and $H\delta$-$D_n(4000)$
diagrams in three different mass bins. In the left-hand panels, we show the BPT diagram at different stellar mass bins. For the low stellar mass sample, most population are star-forming galaxies and most SF activities are also occur in SF region. For higher stellar mass bins, the population moves to composite galaxies and even low-ionization nuclear emission-line region (LINER), but the SF activities still occur between star-forming and composite regions.

In the right-hand panels we see, like \citet{2003MNRAS.341...33K}, that
the strength of the 4000\AA break, tracing evolved stellar populations
($>$ 1 Gyr), is anti-correlated with $H\delta$, tracing the presence of
younger stellar populations (age $<$ 1 Gyr).  Not surprisingly, SF
mostly occurs in galaxies with generally young populations.  For low-mass
galaxies the SF-ing population is representative of the general
population, but for high-mass galaxies the SF-ing galaxies represent a
tail of outliers, as most galaxies are old and quiescent.  Still,
high-mass SF-ing galaxies are older than their low-mass counterparts.
This can be interpreted as evidence for an increasingly prominent
bulge, while SF activity in the disk is similar to that in the disks
of lower-mass galaxies as argued by \citet{2014ApJ...785L..36A}.  However,
it could also signal a decline in SF activity as suggested by
\citet{2014MNRAS.440..889S}.

A detailed exploration of the processes that drive SF evolution is
beyond the scope of this paper, but we hope that the community will
take advantage of the mid-IR based SFR estimates published here to
distinguish further between reddening by dust, age, and reduced SF
activity.

\section{Summary}
\label{sec5}
We revisit the measurement of stellar masses and star-formation rates
for the SDSS spectroscopic galaxy sample after adding 4-band
photometry from WISE for the wavelength range $3-22\mu$m.  We benefit
from this wavelength extension by adopting the latest,
state-of-the-art SED modeling approach
\citep[MAGPHYS][]{2008MNRAS.388.1595D}, which includes the
self-consistent treatment of dust attenuation and emission and a wide
range of star formation histories.  
The resulting SFR esimates are mostly based on
PAH emission and thermal dust radiation, and, therefore, are
complementary to the nebular emission line-derived estimates from
\citep{2004MNRAS.351.1151B}.  In Section \ref{l12_l22_SFR} we provide
new calibrations for the conversion of monochromatic 12$\mu$m and
22$\mu$m luminosities into SFRs.

The new $M_*$ and SFR estimates show the well-established bi-modality
in SF activity.  A first application is the verification of the cruder
color-color separation of SF-ing and quiescent galaxies that is often
used at higher redshifts. We confirm that the $U-V$-$V-J$ (or,
alternatively, $u-r$-$r-z$, as in Figure \ref{swfig_msfr})
color-color separation has a 89\% success rate in identifying SF-ing
and quiescent galaxies.  

We also fit the relation between SFR and $M_*$ (the SF-ing sequence,
or SF-ing Main Sequence).  Our sequence is slightly steeper than
previous studies based on SDSS galaxies, and has a higher
normalization and scatter. We attribute these differences to our new
SFR estimates that now include dust emission and also to some extent
the uncertainty in establishing a purely star-forming sample
(i.e. without contamination from quiescent galaxies).  $L_*$ falaxies
contribute most to the cosmic SF density, and these typically have
optical colors that are in between those of the traditional red
squence and blue cloud (Figure \ref{swfig_mcolor}) as a result of dust
attenuation, not because of reduced star-formation activity.

As a service to the community we make our SDSS+WISE matched photometry
catalog as well as the SED fitting results publicly available.

\acknowledgments

We thank the anonymous referee for useful comments to improve this paper.
Y.-Y. Chang was funded by the IMPRS for Astronomy \& Cosmic Physics at
the University of Heidelberg and the Marie Curie Initial Training
Network ELIXIR of the European Commission under contract
PITN-GA-2008-214227. E. da Cunha acknowledges funding through the ERC grant `Cosmic Dawn'.

Funding for the SDSS and SDSS-II has been provided by the Alfred P. Sloan Foundation, the Participating Institutions, the National Science Foundation, the U.S. Department of Energy, the National Aeronautics and Space Administration, the Japanese Monbukagakusho, the Max Planck Society, and the Higher Education Funding Council for England. The SDSS Web Site is \url{http://www.sdss.org/}.
The SDSS is managed by the Astrophysical Research Consortium for the Participating Institutions. The Participating Institutions are the American Museum of Natural History, Astrophysical Institute Potsdam, University of Basel, University of Cambridge, Case Western Reserve University, University of Chicago, Drexel University, Fermilab, the Institute for Advanced Study, the Japan Participation Group, Johns Hopkins University, the Joint Institute for Nuclear Astrophysics, the Kavli Institute for Particle Astrophysics and Cosmology, the Korean Scientist Group, the Chinese Academy of Sciences (LAMOST), Los Alamos National Laboratory, the Max-Planck-Institute for Astronomy (MPIA), the Max-Planck-Institute for Astrophysics (MPA), New Mexico State University, Ohio State University, University of Pittsburgh, University of Portsmouth, Princeton University, the United States Naval Observatory, and the University of Washington.

This publication makes use of data products from the Wide-field Infrared Survey Explorer, which is a joint project of the University of California, Los Angeles, and the Jet Propulsion Laboratory/California Institute of Technology, funded by the National Aeronautics and Space Administration

The Herschel-ATLAS is a project with Herschel, which is an ESA space observatory with science instruments provided by European-led Principal Investigator consortia and with important participation from NASA. The H-ATLAS website is \url{http://www.h-atlas.org/}.

\begin{table*}
\centering
\caption[]{Input Catalog}
\label{tab_cat_in}
\begin{tabular}{p{3cm}p{1.5cm}p{1.5cm}p{14cm}}
\hline 
\hline
Column Name & Format & Unit  & Column Description \\ 
\hline 
\hline
ID & LONG & - & NYU-VAGC catalog index\\ 
RA & DOUBLE & deg &  J2000 R.A.~[deg] from NYU-VAGC ($r$-band)\\
DEC & DOUBLE & deg & J2000 Dec.~[deg] from NYU-VAGC ($r$-band)\\
REDSHIFT & DOUBLE & - & Redshift from the NYU-VAGC spectroscopic catalog \\
PLATE & LONG & - & SDSS plate from NYU-VAGC \\
MJD & LONG & - & SDSS mjd from NYU-VAGC \\
FIBERID & LONG & - & SDSS fiberid from NYU-VAGC \\
DESIGNATION & STRING & - & ALLWISE designation \\
FLUX0\_U & DOUBLE & Jy & $u$-band Flux before corrections \\
FLUX0\_U\_E & DOUBLE & Jy & Uncertainty of $u$-band flux before corrections \\
FLUX0\_G & DOUBLE & Jy & $g$-band flux before corrections \\
FLUX0\_G\_E & DOUBLE & Jy & Uncertainty of $g$-band flux before corrections \\
FLUX0\_R & DOUBLE & Jy & $r$-band flux before corrections \\
FLUX0\_R\_E & DOUBLE & Jy & Uncertainty of $r$-band flux before corrections \\
FLUX0\_I & DOUBLE & Jy & $i$-band flux before corrections \\
FLUX0\_I\_E & DOUBLE & Jy & Uncertainty of $i$-band flux before corrections \\
FLUX0\_Z & DOUBLE & Jy & $z$-band flux before corrections \\
FLUX0\_Z\_E & DOUBLE & Jy & Uncertainty of $z$-band flux before corrections \\
FLUX0\_W1 & DOUBLE & Jy & $W1$-band flux before corrections \\
FLUX0\_W1\_E & DOUBLE & Jy & Uncertainty $W1$-band of flux before corrections \\
FLUX0\_W2 & DOUBLE & Jy & $W2$-band flux before corrections \\
FLUX0\_W2\_E & DOUBLE & Jy & Uncertainty $W2$-band of flux before corrections \\
FLUX0\_W3 & DOUBLE & Jy & $W3$-band flux before corrections \\
FLUX0\_W3\_E & DOUBLE & Jy & Uncertainty $W3$-band of flux before corrections \\
FLUX0\_W4 & DOUBLE & Jy & $W4$-band flux before corrections \\
FLUX0\_W4\_E & DOUBLE & Jy & Uncertainty $W4$-band of flux before corrections \\
FLUX\_U & DOUBLE & Jy & $u$-band Flux after corrections \\
FLUX\_U\_E & DOUBLE & Jy & Uncertainty of $u$-band flux after corrections \\
FLUX\_G & DOUBLE & Jy & $g$-band Flux after corrections \\
FLUX\_G\_E & DOUBLE & Jy & Uncertainty of $g$-band flux after corrections \\
FLUX\_R & DOUBLE & Jy & $r$-band Flux after corrections \\
FLUX\_R\_E & DOUBLE & Jy & Uncertainty of $r$-band flux after corrections \\
FLUX\_I & DOUBLE & Jy & $i$-band Flux after corrections \\
FLUX\_I\_E & DOUBLE & Jy & Uncertainty of $i$-band flux after corrections \\
FLUX\_Z & DOUBLE & Jy & $z$-band Flux after corrections \\
FLUX\_Z\_E & DOUBLE & Jy & Uncertainty of $z$-band flux after corrections \\
FLUX\_W1 & DOUBLE & Jy & $W1$-band Flux after corrections \\
FLUX\_W1\_E & DOUBLE & Jy & Uncertainty $W1$-band of flux after corrections \\
FLUX\_W2 & DOUBLE & Jy & $W2$-band Flux after corrections \\
FLUX\_W2\_E & DOUBLE & Jy & Uncertainty $W2$-band of flux after corrections \\
FLUX\_W3 & DOUBLE & Jy & $W3$-band Flux after corrections \\
FLUX\_W3\_E & DOUBLE & Jy & Uncertainty $W3$-band of flux after corrections \\
FLUX\_W4 & DOUBLE & Jy & $W4$-band Flux after corrections \\
FLUX\_W4\_E & DOUBLE & Jy & Uncertainty $W4$-band of flux after corrections \\
EXTIN\_U & DOUBLE & $mag$ & Galactic extinction correction for $u$-band\\
EXTIN\_G & DOUBLE & $mag$ & Galactic extinction correction for $g$-band\\
EXTIN\_R & DOUBLE & $mag$ & Galactic extinction correction for $r$-band\\
EXTIN\_I & DOUBLE & $mag$ & Galactic extinction correction for $i$-band\\
EXTIN\_Z & DOUBLE & $mag$ & Galactic extinction correction for $z$-band\\
\hline 
\end{tabular} 
\end{table*}

\begin{table*}
\centering
\caption[]{Output Catalog}
\label{tab_cat_out}
\begin{tabular}{p{3cm}p{1.5cm}p{1.5cm}p{14cm}}
\hline 
\hline
Column Name & Format & Unit  & Column Description \\ 
\hline 
\hline
LMASS\_2\_5 & FLOAT & $\log M_\odot$ & log stellar mass (2.5th percentile) \\
LMASS\_16 & FLOAT & $\log M_\odot$ & log stellar mass (16th percentile) \\
LMASS\_50 & FLOAT & $\log M_\odot$ & log stellar mass (50th percentile) \\
LMASS\_84 & FLOAT & $\log M_\odot$ & log stellar mass (84th percentile) \\
LMASS\_97\_5 & FLOAT & $\log M_\odot$ & log stellar mass (97.5th percentile) \\
LSFR\_2\_5 & FLOAT & $\log M_\odot/yr$ & log SFR (2.5th percentile) \\
LSFR\_16 & FLOAT & $\log M_\odot/yr$ & log SFR (16th percentile) \\
LSFR\_50 & FLOAT & $\log M_\odot/yr$ & log SFR (50th percentile) \\
LSFR\_84 & FLOAT & $\log M_\odot/yr$ & log SFR (84th percentile) \\
LSFR\_97\_5 & FLOAT & $\log M_\odot/yr$ & log SFR (97.5th percentile) \\
LSSFR\_2\_5 & FLOAT & $\log 1/yr$ & log specific SFR (2.5th percentile) \\
LSSFR\_16 & FLOAT & $\log 1/yr$ & log specific SFR (16th percentile) \\
LSSFR\_50 & FLOAT & $\log 1/yr$ & log specific SFR (50th percentile) \\
LSSFR\_84 & FLOAT & $\log 1/yr$ & log specific SFR (84th percentile) \\
LSSFR\_97\_5 & FLOAT & $\log 1/yr$ & log specific SFR (97.5th percentile) \\
LDUST\_2\_5 & FLOAT & $\log L_\odot$ & log dust luminosity (2.5th percentile) \\
LDUST\_16 & FLOAT & $\log L_\odot$ & log dust luminosity (16th percentile) \\
LDUST\_50 & FLOAT & $\log L_\odot$ & log dust luminosity (50th percentile) \\
LDUST\_84 & FLOAT & $\log L_\odot$ & log dust luminosity (84th percentile) \\
LDUST\_97\_5 & FLOAT & $\log L_\odot$ & log dust luminosity (97.5th percentile \\
MU\_2\_5 & FLOAT & - & dust attenuation parameter in \citet{2008MNRAS.388.1595D} (2.5th percentile) \\
MU\_16 & FLOAT & - & dust attenuation parameter in \citet{2008MNRAS.388.1595D} (16th percentile) \\
MU\_50 & FLOAT & - & dust attenuation parameter in \citet{2008MNRAS.388.1595D} (50th percentile) \\
MU\_84 & FLOAT & - & dust attenuation parameter in \citet{2008MNRAS.388.1595D} (84th percentile) \\
MU\_97\_5 & FLOAT & - & dust attenuation parameter in \citet{2008MNRAS.388.1595D} (97.5th percentile) \\
TAUV\_2\_5 & FLOAT & - & dust attenuation parameter in \citet{2008MNRAS.388.1595D} (2.5th percentile) \\
TAUV\_16 & FLOAT & - & dust attenuation parameter in \citet{2008MNRAS.388.1595D} (16th percentile) \\
TAUV\_50 & FLOAT & - & dust attenuation parameter in \citet{2008MNRAS.388.1595D} (50th percentile) \\
TAUV\_84 & FLOAT & - & dust attenuation parameter in \citet{2008MNRAS.388.1595D} (84th percentile) \\
TAUV\_97\_5 & FLOAT & - & dust attenuation parameter in \citet{2008MNRAS.388.1595D} (97.5th percentile) \\
V\_MAX & DOUBLE & $Mpc^3$ & Maximum volume for LMASS\_50\\
LREST\_U & DOUBLE & $\log L_\odot$ & Rest-frame $u$-band Luminosity\\
LREST\_G & DOUBLE & $\log L_\odot$ & Rest-frame $g$-band Luminosity\\
LREST\_R & DOUBLE & $\log L_\odot$ & Rest-frame $r$-band Luminosity\\
LREST\_I & DOUBLE & $\log L_\odot$ & Rest-frame $i$-band Luminosity\\
LREST\_Z & DOUBLE & $\log L_\odot$ & Rest-frame $z$-band Luminosity\\
LREST\_W1 & DOUBLE & $\log L_\odot$ & Rest-frame $W1$-band Luminosity\\
LREST\_W2 & DOUBLE & $\log L_\odot$ & Rest-frame $W2$-band Luminosity\\
LREST\_W3 & DOUBLE & $\log L_\odot$ & Rest-frame $W3$-band Luminosity\\
LREST\_W4 & DOUBLE & $\log L_\odot$ & Rest-frame $W4$-band Luminosity\\
FLAG\_R & INT & - & 1: Simard radius; 2: deVaucouleurs radius; 3: exponential radius; 0: no radius\\
FLAG\_W & INT & - & 1: single optical matched in WISE; 2: $>1$ counterparts within 6"; 0: not matched\\
FLAG\_W1 & INT & - & 1: detected W1 (SNR$>$2); 2: upper limit on W1; 0: no W1 data\\
FLAG\_W2 & INT & - & 1: detected W2 (SNR$>$2); 2: upper limit on W2; 0: no W2 data\\
FLAG\_W3 & INT & - & 1: detected W3 (SNR$>$2); 2: upper limit on W3; 0: no W3 data\\
FLAG\_W4 & INT & - & 1: detected W4 (SNR$>$2); 2: upper limit on W4; 0: no W4 data\\
FLAG\_CHI2 & INT & - & 1: $\chi^2<3$ for best-fit model; 0: $\chi^2>3$ for best-fit model\\
FLAG & INT & - & 1: good fit (FLAG\_R=1; FLAG\_W?=1 or 2; FLAG\_CHI2=1; $z<0.2$); 0: others\\
\hline 
\end{tabular} 
\end{table*}

\bibliographystyle{apj}

\end{document}